\documentclass{article} 

\usepackage[utf8]{inputenc} 
\usepackage[english]{babel} 
\usepackage{amsmath}
\usepackage{amssymb}
\usepackage{txfonts}
\usepackage{mathdots}
\usepackage{graphicx}

\usepackage[authoryear]{natbib}

\begin{document}


\begin{center}
	\Large\textbf{Analytical Solution of Similar Oblate Spheroidal Coordinate System}
\end{center} 

\begin{center}
	\large\textit{Pavel Strunz}\\
	\noindent \textit{(Nuclear Physics Institute of the CAS, 25068 \v{R}e\v{z}, Czech Republic, strunz@ujf.cas.cz)}
\end{center} 
\noindent \\
\noindent Comments: 42 pages, 5 Postscript figures
\noindent \\

\noindent Satisfactory description of gravitational and gravity potentials is needed for a proper modelling of a wide spectrum of physical problems on various size scales, ranging from atmosphere dynamics up to the movements of stars in a galaxy. In certain cases, Similar Oblate Spheroidal (SOS) coordinate system can be of advantage for such modelling tasks, mainly inside or in the vicinity of oblate spheroidal objects (planets, stars, galaxies). Although the solution of the relevant expressions for the SOS system cannot be written in a closed form, it can be derived as analytical expressions -- convergent infinite power series. Explicit analytical expressions for the Cartesian coordinates in terms of the curvilinear Similar Oblate Spheroidal coordinates are derived in the form of infinite power series with generalized binomial coefficients. The corresponding partial derivatives are found in a suitable form, further enabling derivation of the metric scale factors necessary for differential operations. The terms containing derivatives of the metric scale factors in the velocity advection term of the momentum equation in SOS coordinate system are expressed. The Jacobian determinant is derived as well. The presented analytical solution of SOS coordinate system solution is a tool applicable for a broad variety of objects exhibiting density, gravity or gravitation levels resembling similar oblate spheroids. Such objects range from the bodies with small oblateness (the Earth itself on the first place), through elliptical galaxies up to significantly flattened objects like disk galaxies.
\noindent \\

\noindent 
\section{ Introduction}

A proper modelling of gravitational potential as well as gravity potential, and especially the geopotential is of a large importance for weather forecasting and for climate modelling, but also for cosmology. Gravity potential, i.e. a combination of the gravitational potential with a potential of the centrifugal force of a spinning fluid body, is frequently modeled. The equilibrium condition of the rotating spheroidal fluid object derived by \citet{Tod1873} on the basis of Newton theorems \citep{New1729}, and by Clairaut (\citet{Mor1990}, page 45-46 and chapter 5.4) dictates that the constant potential levels (as well as the constant pressure and the constant density levels) in the interior are of a spheroidal shape, moreover fitting with the shape of the spheroidal object itself at the surface. In the case of a homogeneous object (i.e. with a constant density across it), the interior gravity equipotential surfaces are similar oblate spheroids. This is not the case for a spheroidal rotating body in the equilibrium having increasing density with the depth (as e.g. the Earth). For such an object, the deeper equipotential surfaces are increasingly more spherical \citep{Mor1990}. Nevertheless, in a limited range around the surface, a family of similar oblate spheroids can be a satisfactory approximation of the equipotential surfaces, as shown by \citet{Whi2008} for the near-Earth geopotential.

In atmospheric modelling, it is frequent to approximate the geopotential surfaces of the Earth as spheres, and to use spherical polar coordinates to represent the global atmosphere. However Earth's mean surface is more accurately approximated by a spheroid of revolution than by a sphere, and -- therefore -- the geopotential surfaces are better represented as spheroidal surfaces than the spherical ones.

It may be argued that present time ellipticity of the Earth is small enough to use spherical approximation of the geopotential for meteorology and for climate modelling. Nevertheless, it does worth to point out that the Earth's rotation rate is not constant. The Earth's rotation slows by tidal acceleration through gravitational interactions with the Moon \citep{Bar2016} outside the resonant periods. This process gradually increased the length of day \citep{Bar2016,ZW1987} and lowered the centrifugal part of the gravity potential. The former effect had to have a consequence on the temperature \citep{Kra2017} and can probably contribute to the explanation of the Faint Young Sun Paradox \citep{Don1965,Sag1972}, as mentioned by Feulner \citep{Feu2012}. The later effect (lowering the centrifugal part of the gravity potential) had to be accompanied by a shape change of the Earth: originally large equatorial bulge height decreased towards the present value within four billions years. As it proceeded during the long period, the annual accommodation of the Earth shape was in the order of ten micrometres change in the equatorial radius, which probably had negligible effect on tectonics. Nevertheless, the overall change of the Earth's shape was large (tens of kilometres decrease in the equatorial radius). The spherical approximation of the geopotential thus would not be fully sufficient for the detailed pre-historical climate modelling.

Nevertheless, also the present, smaller, magnitude of the ellipticity could have non-negligible effects. Standardly used spherical coordinate system introduces a systematic approximation, and may not be sufficient for proper meteorological modelling as well as for climate modelling on the Earth (see e.g. \citet{Gat2004}).

The use of spherical polar coordinates is widespread in global models, and it is often accompanied by the shallow-atmosphere approximation with no spatial variation of apparent gravity allowed and neglect of various terms in the components of the momentum equation (for discussion of these aspects see \citet{Whi2005}). \citet{Whi2005} showed that inclusion of the latitudinal variation of gravity would be physically inconsistent in a model that assumes the spherical approximation, as spurious sources and sinks of vorticity would occur.

If the Figure of the Earth is not well represented, it could conceivably have significant cumulative effects both in climate-scale integrations and in numerical weather forecasting \citep{Whi2008}. The Earth oblatenes as well as the latitude variation of the magnitude of apparent gravity should be thus better represented.

Suitable models of the Earth's geopotential field provide foundations for modelling Earth's atmosphere and oceans. To facilitate the modelling, appropriate geopotential coordinate system is needed. A coordinate system, in which one coordinate direction is aligned with the local apparent vertical (and the orthogonal coordinate surfaces are geopotentials) provides mathematical as well as conceptual simplification: the magnitude of apparent gravity appears in only one component of the momentum equation, while the two orthogonal components describe the imbalance between horizontal Coriolis and pressure gradient forces that partly determines horizontal particle accelerations \citep{WI2011}.

Several authors described coordinate systems aimed at modelling the gravity potential of the Earth \citep{Gat2004,Whi2008,WI2011,Sta2015,Ben2014,Cha2014}. In the sense of a real analytical orthogonal oblate spheroidal system, which has a potential to fit the real geopotential, the most notable -- although not finalized -- work was done by \citet{Whi2008}. They tested thoroughly a Similar Oblate Spheroidal (SOS) coordinate system which is orthogonal with all advantages \citep{Ben2014} of such type of coordinate systems. In such system, all the orthogonal trajectories to oblate spheroids family (forming the basis of the SOS coordinate system), which are power curves, end up in the common center of the similar oblate spheroids. In the further work \citep{Sta2015}, the authors developed still more general oblate spheroidal coordinate system (titled GREAT). The SOS coordinate system is a limiting case of GREAT. Although the solution of \citet{Sta2015} using a family of oblate spheroids as coordinate surfaces is more general than SOS coordinates, the cost is that the solution cannot be analytical.

\citet{Whi2008} showed that the geopotential is better modeled by similar oblate spheroids in the vicinity of the Earth surface than by confocal oblate spheroid (COS) family \citep{Gat2004}. In both cases, one member of the family fits exactly with the Earth reference ellipsoid. However, the COS family is not a good approximation of the geopotential near the Earth surface (although it is frequently used for this purpose) due to the fact that the gradient of the potential at the surface (determining the acceleration) would be larger at equator than at poles for such a family of confocal spheroids. This contradicts the observations which show that the acceleration is larger at poles ($\mathrm{\approx}$9.83 N/kg) than at the equator ($\mathrm{\approx}$9.78 N/kg) (i.e. by $\mathrm{\approx}$0.5\% when assuming both gravitational and centrifugal component, or by $\mathrm{\approx}$0.2 \% when assuming only gravitational component). This is far larger effect than, e.g., the difference in acceleration on the Earth's surface exerted by the Moon (maximum magnitude $\mathrm{\approx}$1.3$\mathrm{\times}$10${}^{\mathrm{-}6}$ N/kg in that case, see e.g. \citet{Fra2012}) causing tides. The difference in the gravity of the magnitude 0.05 N/kg between equator and poles due to the centrifugal force and the Earth shape is more than 4 orders larger, and should be seriously considered. A correct coordinate system, which can facilitate solving differential equations concerning the Earth's atmosphere, has to take the abovementioned difference in polar and equatorial acceleration into account. Similar oblate spheroids (SOS) coordinate system suggested by \citet{Whi2008} can be a good approximation, as it covers 2/3 of the acceleration difference between the poles and the equator. \\

When assessing gravitational equipotential surfaces of spheroidal objects (from planets to galaxies) in the field of astrophysics, it can be shown that they are approaching spherical shape in the large distances from the object. Therefore, the gravitational equipotential surfaces have to change, when flying from the body towards the infinity, from oblate spheroid surfaces to spherical surfaces in the infinity (see e.g. \citet{Hof2018}). This characteristics is well fulfilled by a confocal oblate spheroid family. In fact, Confocal Oblate Spheroidal (COS) coordinates are frequently used for modelling gravitational potential in the exterior of oblate spheroidal objects (see e.g. \citet{Mor1990}, Chapter 5, or \citet{Hvo2011,Hu2017,Poh2011}).

In the large distances from the object, the gravitational equipotential surfaces cannot be well approximated by a similar oblate spheroids family as the potential differs for them in the polar direction and in the equatorial direction in any distance from the center, even when approaching infinity (as the polar and equatorial semi-axes ratio defining eccentricity is always constant). On the other hand, when going in the reverse direction, i.e. towards the spheroidal object centre, the confocal spheroid surfaces are no more suitable for a description of the gravitational equipotential surfaces of a spheroidal object. When going towards its center, the confocal spheroid surfaces would be becoming increasingly flatter and, eventually, they would change to an infinitely thin flat disc in the equatorial plane. The trajectories orthogonal to that surfaces would then end up not in the center of such system, but on the various points on the focal disc in the equatorial plane. This would contradict to any known gravitation theory.

Indeed, as derived by Maclaurin and Newton (\citet{New1729,Tod1873} or \citet{MMl1958}, section 33), a family of gravitational equipotential surfaces in the interior of a homogeneous body of ellipsoidal shape are mutually similar ellipsoids, although they are not similar to the body ellipsoid itself. (The gravitational equipotential surfaces are ``more spherical'' than the surface of such ellipsoidal body. Therefore, the surface of the ellipsoidal body itself is not a gravitational equipotential surface.) The same has to be true for a spheroid, which is a special case of an ellipsoid. Such internal potential of a homogeneous oblate spheroid was also derived analytically by Gauss and Dirichlet in cylindrical coordinates (see e.g.  \citet{Sch1956}). For completeness, it should be noted that the analytical form of the external potential of a homogeneous oblate spheroid was derived by \citet{Hof2018} in cylindrical coordinates. The continuity of the gravitational equipotential levels inside and outside of the spheroidal body is nicely depicted in the work of \citet{Hvo2011}, their Fig. 2a, which shows the gravitational potential both in the inside and in the outside of an oblate spheroid body. Transition from the equipotential surfaces resembling similar oblate spheroids (near the center) to the surfaces resembling confocal oblate spheroids (far from the center) can be observed.

\citet{Sch1956} derived expressions for potential inside and outside\linebreak non-homogeneous oblate spheroid. He confined the derivation to the case where the eccentricity (sometimes called ellipticity) \textit{e} of all the isodensity levels is constant, which is equivalent to an assumption that all the isodensity levels are similar oblate spheroids sharing the same center and rotation axis. The same approach was used by \citet{Hof2018} for the nested homeoids they report (nevertheless, they focused only on the exterior potential of a spheroid composed of nested homeoids, which is well described using confocal oblate spheroids). On the level of galaxies, it has been shown by \citet{Hof2017} that neither cylindrical geometry nor thin disc geometry are suitable for a description of isodensity and equipotential surfaces. Possibly, a family of similar oblate spheroids could represent better the isodensity levels and/or equipotential surfaces inside galaxies. In fact, similar oblate spheroidal isodensity levels were used in the follow-up work \citet{Cri2020}.

Sumarizing the previous paragraphs, the gravitational potential outside oblate\linebreak spheroid objects can be well modeled with a help of Confocal Oblate Spheroidal (COS) coordinates, while the gravitational potential in the interior of oblate spheroid objects could be possibly advantageously modeled within Similar Oblate Spheroidal (SOS) coordinate system. If the gravitational potential levels fit with similar oblate spheroids family, the use of SOS system would mean transformation of the potential equation to one dimensional problem (i.e. depending on one coordinate of the SOS system only). It would lead to an essential simplification for solution of differential equations. \\

A question might arise on the usefulness of the SOS coordinate system in the view of the fact that an isodensity levels and potential stratification by similar oblate spheroidal surfaces is impossible, as can be found e.g. in \citet{Mor1990}, after \citet{Wav1932}. Nevertheless, apart from the application for simple demonstrative examples like potential levels inside a homogeneous spheroid, two additional points has to be mentioned. First, similar oblate spheroids can be a good approximation to be used in a limited range in the vicinity of the real planet surface, as was suggested by \citet{Whi2008}. Secondly the impossibility of the gravity potential stratification by spheroidal levels was proved only for spinning Earth-like objects which have a constant angular velocity everywhere inside the object. This is, nevertheless, not the case for galaxies in which the rotational curves indicate that they are certainly not fully spinning (although a combination of spinning with star orbiting may take place). Further, another use of the SOS coordinates (not discussed in this introduction) may arise in other fields of physics.

The SOS coordinate system can be thus very helpful for modelling of density levels and/or gravitational or gravity potential in some regions inside or in the vicinity of oblate spheroidal objects (Earth, planets, stars, galaxies) providing that a transformation from SOS to Cartesian coordinates as well as the metric scale factors can be written in an analytical form. Such a tool is still missing in the field of gravity or gravitational potential modelling.

\citet{Whi2008} solved analytically a large part of the task. Nevertheless, the last step -- evaluation of the metric scale factors of the SOS coordinate system -- was not finalized. The authors \citep{Whi2008} stated that the relevant formulas (connecting Cartesian coordinates with the SOS coordinates) cannot be solved analytically in cases that are of interest. This is, nevertheless, not a fully correct statement if we assume also infinite series as an analytic solution, as is the usual consensus in mathematics. In mathematics, an analytic function is a function that is locally given by a convergent power series. Although the solution of the relevant expressions for SOS coordinate system really cannot be written in a closed form, it can be written as analytical expression -- convergent infinite power series.

This text has the aim to finalize the SOS coordinate system description, to derive its metric scale factors, and indicate their use in differential operations. The organization of the text is the following. First, the present state of the derivation of the SOS coordinate system is presented (Section 2). The procedure how to express analytically the individual Cartesian coordinates in dependence on SOS coordinates with the use of infinite power series is shown in Sections 3. The transformation experssions are derived in Section 4. This section also includes discussion on convergence of the used power series at the border line between two distinct regions. Section 5 shows simple numerical determination of the Cartesian coordinates from the SOS coordinates. Section 6 shows the transformation for a particular SOS system at the border line between two distinct regions. Section 7 shows the analytical transformation of the SOS coordinates to the Cartesian coordinates and with the inverse transformation. The partial derivatives and the scale factors are derived in Sections 8 and 9. The Jacobian determinant, frequently employed in differential operations, is derived in Section 10. Section 11 deals with the derivation of components of the advection term in the momentum equation. The gravitational potential and force in the interior of a homogeneous oblate spheroid in SOS coordinates is shown in Section 12. A code written for numerical calculations and tests is mentioned in Section 13. Due to the fact that some derivations with an extensive use of combinatorial identities are lengthy, a large part of the derivations of the formulas can be found in Supplements to this article.

\noindent 
\section{ Similar oblate spheroidal (SOS) coordinates -- the state of the art and coordinates labeling}

Setting up an orthogonal curvilinear coordinate system based on similar oblate\linebreak spheroids requires the spheroids to be labelled, and the orthogonal surfaces to them to be identified and labelled. This was done by \citet{Whi2008}. They labelled ellipses (which generate spheroids of the SOS system) conventionally with their equatorial radius \textit{R}. A family of similar ellipses having major semi-axis \textit{R} may be specified analytically as
\begin{equation} \label{GrindEQ__1_} 
	{x}^{2} +\left(1+\mu \right){z}^{2} =\ {R}^{2} .     
\end{equation} 
The parameter \textit{$\mu$} characterizes the whole family of similar ellipses having the same eccentricity (and various major semi-axes). The parameter \textit{$\mu$} also characterizes the whole family of the similar oblate spheroids generated by rotating the ellipses similar to \eqref{GrindEQ__1_} in 3D around the minor axis. The minor and major semi-axes of each member have the ratio (1+\textit{$\mu$})${}^{\mathrm{-}}$${}^{1/2}$, and \textit{$\mu$} is connected with the eccentricity \textit{e} and flattening \textit{f} by the relations
\begin{equation} \label{GrindEQ__2_} 
	{e =\sqrt{\frac{\mu }{1+\mu }}}\ ,\ \ 1+\mu ={\frac{1}{1-{e}^{2}}}\ ,\ \ \mu ={\frac{{e}^{2}}{1-{e}^{2}} =\frac{1}{{\left(1-f\right)}^{2}}-1\ ,\ \ \ f=1-\frac{1}{\sqrt{1+\mu }}\ } .  
\end{equation} 
It has to be noted that the constant parameter \textit{$\mu$} also means the constant eccentricity \textit{e}. A family of similar oblate spheroids (although explicitly not titled as such) was in fact used e.g. by \citet{Sch1956} to describe his non-homogeneous oblate spheroid isodensity levels.

Longitude angle $\mathrm{\lambda}$ was, naturally, chosen by \citet{Whi2008} as the third orthogonal coordinate of the orthogonal coordinate system based on the similar oblate spheroids. The associated coordinate surfaces are then semi-infinite planes.

What remains to determine is the second (meridional) orthogonal coordinate of the SOS system. The curves orthogonal to the similar ellipses that generate the similar spheroids were determined analytically by \citet{Whi2008}. The orthogonal trajectories of the family \eqref{GrindEQ__1_} are the power curves
\begin{equation} \label{GrindEQ__3_} 
{\ z =\ }\ {Dx}^{\mathrm{1+}\mathrm{\mu }}\  .      
\end{equation} 
The associated meridional coordinate surfaces are generated by rotating them around the rotation axis of the spheroids. The labelling was carried out by \citet{Whi2008} in the way that the meridional coordinate $\mathrm{\varphi}$ is connected with the above parameter \textit{D} by the following relation:
\begin{equation} \label{GrindEQ__4_} 
D={\frac{D_{\mathrm{WSW}}}{a^{\mu }_0} =\ }\frac{\mathrm{1}}{{\left(\mathrm{1+}\mathrm{\mu }\right)}^{\mathrm{1/2}}a^{\mu }_0}\mathrm{\ }{\mathrm{tan}}^{\mathrm{1+}\mathrm{\mu }}\mathrm{\varphi }\mathrm{\ \ \ \ \ \ \ for\ \ }D\mathrm{,\ \ }\mathrm{\varphi }\mathrm{\ }\mathrm{\ge }\mathrm{0}\  
\end{equation} 
where \textit{D}${}_{WSW}$ is the original proportionality parameter of \citet{Whi2008} who used scaled coordinates \textit{x} and \textit{z} (scaled by the equatorial radius of the Earth \textit{a}${}_{0}$) unlike in this manuscript where non-scaled coordinates are used.

Differently from \citet{Whi2008}, a meridional coordinate \textit{$\nu$} is employed here for which the following relation to \textit{D} holds: 
\begin{equation} \label{GrindEQ__5_} 
{D =\ }\frac{\mathrm{1}}{{\left(\mathrm{1+}\mathrm{\mu }\right)}^{{\mathrm{1}}/{\mathrm{2}}}\mathrm{\ }R^{\mu }_0}\mathrm{\ }\frac{{\mathrm{sin} \mathrm{\nu }\ }}{{\mathrm{cos}}^{\mathrm{1+}\mathrm{\mu }}\mathrm{\ }\mathrm{\nu }}\mathrm{\ \ } 
\end{equation} 
Using this labelling, the meridional coordinate \textit{$\nu$} is also equivalent to the coordinate used for standard parametric equation of ellipse with \textit{R}${}_{0}$ major semi-axis. At this ellipse, 
\begin{equation} \label{GrindEQ__6_} 
x=R_0{\mathrm{cos} \mathrm{\nu }\ }\   .     
\end{equation} 
and
\begin{equation} \label{GrindEQ__7_} 
z=\frac{R_0}{\sqrt{\mathrm{1+}\mu }}{\mathrm{sin} \mathrm{\nu }\ }  .     
\end{equation} 
This reference ellipse generates the reference spheroid with the equatorial radius \textit{R}${}_{0}$. Therefore, \textit{$\nu$} is equivalent to the so called parametric latitude. Further, this type of labelling of the direction, in which the latitude changes (meridional coordinate), seems to be the most convenient labelling choice if one intends to have a connection between the confocal oblate spheroidal (COS) coordinate system and the similar oblate spheroidal (SOS) systems on one particular oblate spheroid surface with the equatorial radius \textit{R}${}_{0}$. A point at that spheroid surface with the equatorial radius \textit{R}${}_{0}$ has the same value for the meridional coordinate \textit{$\nu$} of the SOS system and for the meridional coordinate \textit{$\nu$}${}_{C}$ of the COS system. Nevertheless, the possibility of fitting SOS and COS on a particular spheroidal surface will not be further discussed in this paper, as it would complicate the explanation of the SOS coordinates unnecessarily.

\begin{figure}
	\includegraphics[width=\linewidth]{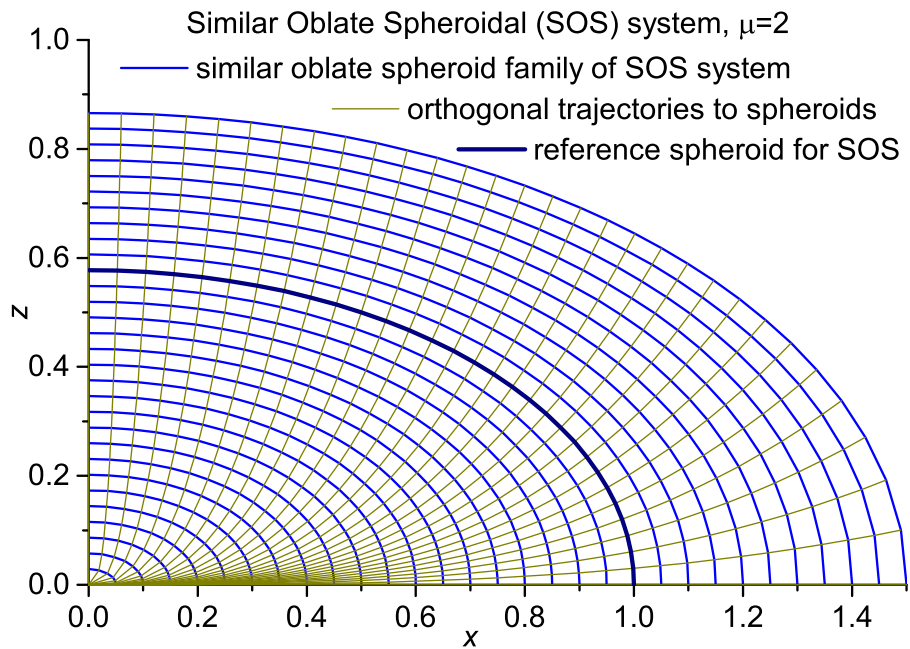}
	\caption{One quadrant composed of the SOS coordinate system (with \textit{µ}=2) lines around the reference spheroid with the equatorial radius \textit{R}${}_{0}$=1. Both constant-\textit{R} coordinate surfaces (spheroids, here represented by their sections by \textit{x}-\textit{z} plane) and orthogonal trajectories to them with a constant value of $\mathrm{\nu}$ are displayed. The angular spacing of the parametric angle $\mathrm{\nu}$ is 3$\mathrm{{}^\circ}$.}
	\label{fig:Fig1}
\end{figure}

An example of the SOS coordinate system with \textit{µ}=2 around a reference spheroid with the equatorial radius \textit{R}${}_{0}$=1 (i.e. with the polar radius equal to ${1}/{\sqrt{1+2}}\cong \mathrm{0.57735}$) is shown in \textbf{Figure \ref{fig:Fig1}}.

In this paper, we will use the meridional coordinate \textit{$\nu$} of the orthogonal trajectories \eqref{GrindEQ__3_} defined according to the formula \eqref{GrindEQ__5_} in the derivation of SOS-relevant formulas (the relation of this parametric latitude to the geocentric latitude and to the geographic latitude can be found in \citet{Gat2004}, Eqs. 38a, 38b, or in \citet{Sta2015}, Table 1). Nevertheless, a transformation of the resulting formulas to the meridional coordinate $\mathrm{\varphi}$ defined by \eqref{GrindEQ__4_} can be easily obtained.

\noindent 
\section{ Problem of explicit expressions for coordinates and starting point of its solution}

If explicit expressions existed for \textit{x}, \textit{y} and \textit{z} in terms of \textit{$\lambda$}, \textit{$\nu$} and \textit{R} of the SOS system, then it would be possible to transform the coordinate systems between themselves and to calculate metric scale factors of the SOS coordinate system. However, this is not possible to carry out in a straightforward way. When trying to separate and to express the variables \textit{x} and \textit{z} (in \textit{x}-\textit{z} plane) explicitly only in terms of equatorial radius \textit{R} and meridional coordinate \textit{$\nu$} (i.e. the coordinates of the SOS system), which is necessary for calculating derivatives and, consequently, the metric scale factors \textit{h${}_{\lambda }$, h${}_{R}$} and \textit{h${}_{\nu }$}, one arrives \citep{Whi2008} $\mathrm{-}$ using \eqref{GrindEQ__1_} and \eqref{GrindEQ__3_} $\mathrm{-}$ at the parametric equation
\begin{equation} \label{GrindEQ__8_} 
{{\left(\frac{\mathrm{1}}{\mathrm{\ }R^{\mu }_0}\frac{{\mathrm{\ \ \ sin}}\mathrm{\nu }}{{\mathrm{\ \ \ cos}}^{\mathrm{1+}\mathrm{\mu }}\mathrm{\ }\mathrm{\nu }}\right)}^{\mathrm{2}}{\left(x^{\mathrm{2}}\right)}^{\mathrm{1+}\mu }\mathrm{+}x^{\mathrm{2}} =\ }\ R^{\mathrm{2}}\  .     
\end{equation} 
This is, except special cases of \textit{µ}, an equation which cannot be solved for \textit{x} in a closed form. Nevertheless, it can be solved in the form of infinite power series using Lagrange Inversion Formula \citep{Lag1770}. A solution of equation of a similar type (\textit{x}${}^{N}$ $\mathrm{-}$ \textit{x} + \textit{t} = 0, \textit{N}=2,~3,~4~...) was found by  \citet{Gla1994}. Still much earlier (1925), and more generally, this problem was solved by \citet{Pol1925}, who determined a combinatorial identity connected with the equation similar to \eqref{GrindEQ__8_}, particularly
\begin{equation} \label{GrindEQ__9_} 
\mathrm{-}uy^b\mathrm{+}y\mathrm{=1}\  .     
\end{equation} 
This combinatorial identity is
\begin{equation} \label{GrindEQ__10_} 
y^a=\sum^{\mathrm{\infty }}_{k\mathrm{=0}}{\frac{a}{a\mathrm{+}bk}\left(\genfrac{}{}{0pt}{}{a\mathrm{+}bk}{k}\right)u^k}\  
\end{equation} 
and is listed in \citet{Gou1972} collection of combinatorial identities under the number 1.121 (another, perhaps more easily accessible reference where mentioned, is \citet{Gou1956}). It includes generalized binomial coefficients. There is no restriction on the exponent \textit{b} in \eqref{GrindEQ__9_} which means (compare with exponent in (8)) that the solution can be obviously used for oblate spheroids even with the small values of \textit{$\mu$} valid for the Earth reference ellipsoid (\textit{$\mu$} = 0.006739496742) as well as for large \textit{$\mu$} values, and probably even for prolate spheroids ($\mathrm{-}$1$\mathrm{<}$\textit{$\mu$$<$0}).

There is also a second combinatorial identity of the same authors listed in \citet{Gou1972} collection of combinatorial identities under the number 1.120, and mentioned in \citet{Gou1956}, connected with the solution of the equation \eqref{GrindEQ__9_}:
\begin{equation} \label{GrindEQ__11_} 
\frac{y^{a+1}}{\left(\mathrm{1-}b\right)y\mathrm{+}b}=\sum^{\mathrm{\infty }}_{k\mathrm{=0}}{\left(\genfrac{}{}{0pt}{}{a\mathrm{+}bk}{k}\right)u^k}\  
\end{equation} 
This identity, especially for the case when \textit{a}=$\mathrm{-}$1, leads also to a relatively simple analytical solution of \eqref{GrindEQ__9_} equation, and will be also used in this article. Moreover, \citet{Gou1956} gave the convergence limit for the infinite power series in \eqref{GrindEQ__10_} and \eqref{GrindEQ__11_}:
\begin{equation} \label{GrindEQ__12_} 
\left|u\right|<\left|\frac{{\left(b\mathrm{-}\mathrm{1}\right)}^{b\mathrm{-}\mathrm{1}}}{{\left.b\right.}^b}\right| 
\end{equation} 
(see also \citet{Gou1972}). It is declared elsewhere \citep{LIT2020} that the series of \eqref{GrindEQ__10_} type converge not only for $\left|u\right|$ smaller than, but also for $\left|u\right|$ equal to the quantity on the right side of \eqref{GrindEQ__12_}. It should be noted that the convergence does not depend on the parameter \textit{a}, but solely on the exponent \textit{b} of the equation to be solved.

For completeness, we report also another solution resembling the equation \eqref{GrindEQ__8_} to be solved in our case. \citet{Bag2014} derived for the equation
\begin{equation} \label{GrindEQ__13_} 
u_{\mathrm{B}}qy^p_{\mathrm{B}}\mathrm{+}y^q_{\mathrm{B}}\mathrm{=1}\  
\end{equation} 
a solution in the form
\begin{equation} \label{GrindEQ__14_} 
y^n_{\mathrm{B}}=\frac{n}{q}\sum^{\mathrm{\infty }}_{k\mathrm{=0}}{\frac{\mathrm{\Gamma }\left(\left\{n\mathrm{+}pk\right\}\mathrm{/}q\right){\left(\mathrm{-}qu_{\mathrm{B}}\right)}^k}{\mathrm{\Gamma }\left(\left\{n\mathrm{+}pk\right\}\mathrm{/}q\mathrm{-}k\mathrm{+1}\right)\mathrm{\ }k\mathrm{!}}}\ \ \ ,\ \ \ n=1,\ 2,\ 3,\ \dots  ,   
\end{equation} 
where the symbol $\Gamma$ denotes Euler's Gamma function. The equation \eqref{GrindEQ__13_} can be transformed to \citet{Pol1925} equation \eqref{GrindEQ__9_} by substitution $y=y^q_{\mathrm{B}}$, \textit{b}=\textit{p}/\textit{q} and --\textit{u }= \textit{u}${}_{B}$\textit{q}. It can be shown using binomial identities (see the introductory notes in \textbf{Supplement A}) that the solution \eqref{GrindEQ__14_} is equivalent with the solution \eqref{GrindEQ__10_}, except that \citet{Bag2014} restricts himself to the solution exponents \textit{n}=1,~2,~3~{\dots}, whereas \citet{Pol1925} enabled solution for any real exponent \textit{a}.

The \citet{Pol1925} solution \eqref{GrindEQ__10_} will be used in what follows for searching the explicit expressions for \textit{x} and \textit{z} in terms of \textit{$\nu$} and \textit{R}. This is in fact giving only 2D solution in \textit{x-z} plane, i.e. describes an ellipse. Nevertheless, the 3D oblate spheroid solution (3D Cartesian coordinates \textit{x}${}_{3D}$, \textit{y}${}_{3D}$, \textit{z}${}_{3D}$ in terms of \textit{$\lambda$}, \textit{$\nu$} and \textit{R}) can be then simply obtained by a rotation of the resulting ellips1 e around the minor axis (\textit{z}-axis), i.e. 
\begin{equation} \label{GrindEQ__15_} 
x_{\mathrm{3D}}\ \mathrm{=\ }x\left(\nu ,R\right)\mathrm{cos\ }\lambda ,\ \ \ \ \ \ \ \ \ \ \ \ \ \ \ \ \ y_{\mathrm{3D}}\mathrm{=\ }x\left(\nu ,R\right)\mathrm{sin\ }\lambda ,\ \ \ \ \ \ \ \ \ \ \ \ \ \ \ \ \ \ z_{\mathrm{3D}}\ \mathrm{=\ }z(\nu ,R)\ .  
\end{equation} 
After \textit{x}(\textit{$\nu$},\textit{R}) and \textit{z}(\textit{$\nu$},\textit{R}) are found, the scale factors \textit{h${}_{\lambda }$, h${}_{R}$} and \textit{h${}_{\nu }$} will be calculated using the formulas \citep{Whi2008}
\begin{equation} \label{GrindEQ__16_} 
h_{\lambda }\mathrm{=}\left|x\left(\nu ,R\right)\right|\ ,\ \ \ h_{\nu }\mathrm{=}\sqrt{{\left(\frac{\partial x\left(\nu ,R\right)}{\partial \nu }\right)}^2+{\left(\frac{\partial z\left(\nu ,R\right)}{\partial \nu }\right)}^2}\mathrm{\ },\ \ \ h_R\mathrm{=}\sqrt{{\left(\frac{\partial x\left(\nu ,R\right)}{\partial R}\right)}^2+{\left(\frac{\partial z\left(\nu ,R\right)}{\partial R}\right)}^2}\ .  
\end{equation} 

Further, as \textit{R} and \textit{$\nu$} are orthogonal coordinates, the following relation, which will appear useful in later derivations of components of the advection formula, holds \citep{Gil1954,Whi2008}:
\begin{equation} \label{GrindEQ__17_} 
\frac{\partial x\left(\nu ,R\right)}{\partial \nu }\frac{\partial x\left(\nu ,R\right)}{\partial R}+\frac{\partial z\left(\nu ,R\right)}{\partial \nu }\frac{\partial z\left(\nu ,R\right)}{\partial R}=0\ .    
\end{equation} 

To modify it to a useful form, we can separate derivatives of \textit{z}(\textit{R},\textit{$\nu$}) from \eqref{GrindEQ__16_}, which -- in the first quadrant where both derivatives of \textit{z}(\textit{R},\textit{$\nu$}) are positive -- leads to
\begin{equation} \label{GrindEQ__18_} 
\frac{\partial z\left(\nu ,R\right)}{\partial \nu }=\sqrt{h^2_{\nu }-{\left(\frac{\partial x\left(\nu ,R\right)}{\partial \nu }\right)}^2}\mathrm{\ },\ \ \ \ \ \ \ \ \ \ \ \ \ \ \ \mathrm{\ }\frac{\partial z\left(\nu ,R\right)}{\partial R}=\sqrt{h^2_R-{\left(\frac{\partial x\left(\nu ,R\right)}{\partial R}\right)}^2}\ .  
\end{equation} 

Note, that the derivative of \textit{x}(\textit{R},\textit{$\nu$}) with respect to \textit{R} is positive while the derivative of \textit{x}(\textit{R},\textit{$\nu$}) with respect to \textit{$\nu$} is negative in the first quadrant. When eliminating derivatives of \textit{z}(\textit{R},\textit{$\nu$})) in \eqref{GrindEQ__17_} by means of \eqref{GrindEQ__18_}, and when employing the fact that \textit{x}(\textit{R},\textit{$\nu$}) is positive in the first quadrant (and thus \textit{h}${}_{\mathrm{\lambda }}$={\textbar}\textit{x}(\textit{R},\textit{$\nu$}){\textbar}=\textit{x}(\textit{R},\textit{$\nu$}), see \eqref{GrindEQ__16_}), we arrive at the formula
\begin{equation} \label{GrindEQ__19_} 
\frac{\partial x\left(\nu ,R\right)}{\partial \nu }\frac{\partial x\left(\nu ,R\right)}{\partial R}+\frac{\partial z\left(\nu ,R\right)}{\partial \nu }\frac{\partial z\left(\nu ,R\right)}{\partial R}=\frac{\partial h_{\lambda }}{\partial \nu }\frac{\partial h_{\lambda }}{\partial R}+\sqrt{h^2_{\nu }-{\left(\frac{\partial h_{\lambda }}{\partial \nu }\right)}^2}\sqrt{h^2_R-{\left(\frac{\partial h_{\lambda }}{\partial R}\right)}^2}=0\ .  
\end{equation} 

Note, that ${\partial h_{\lambda }}/{\partial R}$ is positive while ${\partial h_{\lambda }}/{\partial \nu }$ is negative in the first quadrant. By a subsequent algebraic manipulation, we receive 
\begin{equation} \label{GrindEQ__20_} 
{\left(\frac{\partial h_{\lambda }}{\partial \nu }\right)}^2h^2_R=\ h^2_{\nu }\left(h^2_R-{\left(\frac{\partial h_{\lambda }}{\partial R}\right)}^2\right)\ \ \ \ \ \ \ \ \ {{\stackrel{}{\Rightarrow}\ \ \ }}\ \ \ \ -\frac{\partial h_{\lambda }}{\partial \nu }h_R=\ h_{\nu }\sqrt{h^2_R-{\left(\frac{\partial h_{\lambda }}{\partial R}\right)}^2}\ \  ,  
\end{equation} 
which relation will turn useful in later derivations performed in the first quadrant.

\noindent 
\section{ Expressions for \textit{x} and \textit{z} in dependence on coordinates of the SOS system using power series with generalized binomial coefficients }

If we rearrange the formula \eqref{GrindEQ__8_} in the way that
\begin{equation} \label{GrindEQ__21_} 
{{\left({\left(\frac{R}{\mathrm{\ }R_0}\right)}^{\mu }\frac{{\mathrm{\ \ \ sin}}\mathrm{\nu }}{{\mathrm{\ \ \ cos}}^{\mathrm{1+}\mathrm{\mu }}\mathrm{\ }\mathrm{\nu }}\right)}^{\mathrm{2}} {\left(\frac{x}{R}\right)}^{\mathrm{2+2}\mu }+{\left(\frac{x}{R}\right)}^{\mathrm{2}}=\ }\ 1\  
\end{equation} 
then -- with substitutions \textit{s}= \textit{x}/\textit{R}, and defining (with help of (5)) a new parameter
\begin{equation} \label{GrindEQ__22_} 
W\equiv W(R,\nu )={\left(\frac{R}{\mathrm{\ }R_0}\right)}^{\mu }\frac{{\mathrm{\ \ \ sin}}\mathrm{\nu }}{{\mathrm{\ \ \ cos}}^{\mathrm{1+}\mathrm{\mu }}\mathrm{\ }\mathrm{\nu }}={\mathrm{(1+}\mu \mathrm{)}}^{{\mathrm{1}}/{\mathrm{2}}}\mathrm{\ }D\mathrm{\ }R^{\textrm{µ}} 
\end{equation} 
-- we arrive at the equation
\begin{equation} \label{GrindEQ__23_} 
{W^{\mathrm{2}} {\left(s^{\mathrm{2}}\right)}^{\mathrm{1+}\mu }+s^{\mathrm{2}}=\ }\ 1\  ,     
\end{equation} 
which resembles the \citet{Pol1925} equation \eqref{GrindEQ__9_} with \textit{u}= --\textit{W}${}^{2}$, \textit{y}=\textit{s}${}^{2}$ and \textit{b}=1+\textit{µ}. According to \eqref{GrindEQ__10_}, its solutions are
\begin{equation} \label{GrindEQ__24_} 
s\mathrm{=}\frac{x}{R}\mathrm{=}{\left(s^2\right)}^{{{\frac{1}{2}}}}=\sum^{\mathrm{\infty }}_{k\mathrm{=0}}{\frac{{{\frac{1}{2}}}}{{{\frac{1}{2}}}\mathrm{+}\left(1+\mu \right)k}\left(\genfrac{}{}{0pt}{}{{{\frac{1}{2}}}\mathrm{+}\left(1+\mu \right)k}{k}\right)}{\left(\mathrm{-}W^{\mathrm{2}}\right)}^k\ \  
\end{equation} 
and
\begin{equation} \label{GrindEQ__25_} 
s^{1+\mu }\mathrm{=}{\left(\frac{x}{R}\right)}^{1+\mu }={\left(s^2\right)}^{{{\frac{1+\mu }{2}}}}=\sum^{\mathrm{\infty }}_{k\mathrm{=0}}{\frac{{{\frac{1+\mu }{2}}}}{{{\frac{1+\mu }{2}}}\mathrm{+}\left(1+\mu \right)k}\left(\genfrac{}{}{0pt}{}{{{\frac{1+\mu }{2}}}\mathrm{+}\left(1+\mu \right)k}{k}\right)}{\left(\mathrm{-}W^{\mathrm{2}}\right)}^k\ .  
\end{equation} 

In what follows, we will restrict the calculation to the meridional angles positioned only in the first quadrant of the Cartesian coordinate system, thus the generating ellipse is calculated only for $\nu \in \left\langle 0\right.,\left.{{\frac{\pi }{2}}}\right\rangle \ $. Due to the symmetry, the solution in the other quadrants can be easily obtained. With this restriction, the parameter \textit{W} is always non-negative.

As (see (22)) \textit{W}= (1+\textit{$\mu$})${}^{1\mathrm{/}2}$ \textit{D} \textit{R${}^{\textrm{µ}}$}, we can simply use Eq. \eqref{GrindEQ__3_} and \eqref{GrindEQ__25_} for the calculation of the coordinate \textit{z}:

\begin{equation} \label{GrindEQ__26_} 
	\begin{aligned} 
&z={Dx}^{\mathrm{1+}\mathrm{\mu }}=\frac{W}{{\left(1+\mu \right)}^{1/2}R^{\mu }}R^{1+\mu }s^{1+\mu }  \\
&=\frac{W}{{\left(1+\mu \right)}^{1/2}}R\sum^{\mathrm{\infty }}_{k\mathrm{=0}}{\frac{{{\frac{1+\mu }{2}}}}{{{\frac{1+\mu }{2}}}\mathrm{+}\left(1+\mu \right)k}\left(\genfrac{}{}{0pt}{}{{{\frac{1+\mu }{2}}}\mathrm{+}\left(1+\mu \right)k}{k}\right)}{\left(\mathrm{-}W^{\mathrm{2}}\right)}^k .  
	\end{aligned} 
\end{equation} 
The coordinate \textit{x} can be retrieved from \eqref{GrindEQ__24_}:
\begin{equation} \label{GrindEQ__27_} 
x=R\sum^{\mathrm{\infty }}_{k\mathrm{=0}}{\frac{{{\frac{1}{2}}}}{{{\frac{1}{2}}}\mathrm{+}\left(1+\mu \right)k}\left(\genfrac{}{}{0pt}{}{{{\frac{1}{2}}}\mathrm{+}\left(1+\mu \right)k}{k}\right)}{\left(\mathrm{-}W^{\mathrm{2}}\right)}^k\ \ \  .   
\end{equation} 
With this, we have expressed \textit{x} and \textit{z} explicitly in terms of \textit{$\nu$} and \textit{R} only, through the parameter \textit{W}(\textit{R},\textit{$\nu$}) defined in \eqref{GrindEQ__22_}. It does worth to note, that the parameter \textit{W} can be equally well expressed (with the the help of (4) and \eqref{GrindEQ__22_}) also in terms of \citet{Whi2008} coordinate $\mathrm{\varphi}$: 
\begin{equation} \label{GrindEQ__28_} 
W(R,\mathrm{\varphi })=\mathrm{\ }\frac{R^{\textrm{µ}}}{a^{\mu }_0}\mathrm{\ }{\mathrm{tan}}^{\mathrm{1+}\mathrm{\mu }}\mathrm{\varphi }\ .    
\end{equation} 

The convergence of the power series in \eqref{GrindEQ__26_} and \eqref{GrindEQ__27_} has to be carefully considered. We can use the convergence limit \eqref{GrindEQ__12_} for this purpose. After the substitution (1+\textit{$\mu$}) for \textit{b} and --\textit{W}${}^{2}$ for \textit{u} in \eqref{GrindEQ__12_}, we arrive at 
\begin{equation} \label{GrindEQ__29_} 
\left|-W^2\right|<\left|\frac{{\mu }^{\mu }}{{\left.\left(\mathrm{1+}\mu \right)\right.}^{\mathrm{1+}\mu }}\right|\mathrm{\ \ \ \ \ \ \ \ \ }{{\stackrel{}{\Rightarrow}}}\mathrm{\ \ \ \ \ \ \ }\left|W\right|<\sqrt{\frac{{\mu }^{\mu }}{{\left(\mathrm{1+}\mu \right)}^{\mathrm{1+}\mu }}}\mathrm{=}\sqrt{\frac{\mathrm{1}}{\left.\mathrm{1+}\mu \right.}{\left(\frac{\mu }{\mathrm{1+}\mu }\right)}^{\mu }}\mathrm{\ }\ ,  
\end{equation} 
where the rightmost expression is of advantage for a numerical calculation for large \textit{µ}. Indeed, this convergence limit works, as proved numerically. The obtained limit \eqref{GrindEQ__29_} shows that there is a maximum of the parameter \textit{W} (defined by (22)) for which the formulas \eqref{GrindEQ__26_} and \eqref{GrindEQ__27_} can be employed. For larger values of \textit{W}, another solution of \eqref{GrindEQ__8_} has to be searched for. Depending on \textit{µ}, the limit for \textit{W} lies between 1 (for \textit{µ}=0) and 0 (for \textit{µ} approaching infinity). Therefore, the solutions \eqref{GrindEQ__26_} and \eqref{GrindEQ__27_} can be used only for small meridional coordinate \textit{$\nu$}, nearer to the equatorial plane (for which \textit{$\nu$=0}), while it cannot be used (as the series does not converge) for the large absolute meridional coordinate \textit{$\nu$} nearer to the pole (for which \textit{$\nu$=$\pi$/2}). In what follows, these two regions are called ``small-\textit{$\nu$} region'' and ``large-\textit{$\nu$} region'', respectively.

For covering also the large-\textit{$\nu$} region, the parametric equation \eqref{GrindEQ__23_} can be -- with a help of the substitution
\begin{equation} \label{GrindEQ__30_} 
t^{{{\frac{2}{\mathrm{1+}\mathrm{\mu }}}}}{= {\left(W^{\mathrm{2}}\right)}^{{{\frac{\mathrm{1}}{\mathrm{1+}\mathrm{\mu }}}}}\mathrm{\ }s^{\mathrm{2}}\ }={\left(W^{\mathrm{2}}\right)}^{{{\frac{\mathrm{1}}{\mathrm{1+}\mathrm{\mu }}}}}\mathrm{\ }{\left(\frac{x}{R}\right)}^{\mathrm{2}} 
\end{equation} 
-- rewritten to the form
\begin{equation} \label{GrindEQ__31_} 
{{W^{{{\frac{\mathrm{2}}{\mathrm{1+}\mathrm{\mu }}}}}\mathrm{\ }\left(t^{\mathrm{2}}\right)}^{\frac{1}{\mathrm{1+}\mathrm{\mu }}} \ +\ t^{\mathrm{2}}=\ }\ 1 ,     
\end{equation} 
which resembles the \citet{Pol1925} equation \eqref{GrindEQ__9_} with \textit{u}=--${\left(W^{\mathrm{2}}\right)}^{{{\frac{\mathrm{1}}{\mathrm{1+}\mathrm{\mu }}}}}$, \textit{y}=\textit{t}${}^{2}$ and \textit{b}=1/(1+\textit{µ}). According to \eqref{GrindEQ__10_}, its solutions are
\begin{equation} \label{GrindEQ__32_} 
t^{{{\frac{1}{\mathrm{1+}\mathrm{\mu }}}}}\mathrm{=}W^{{{\frac{\mathrm{1}}{\mathrm{1+}\mathrm{\mu }}}}}\mathrm{\ }\frac{x}{R}\mathrm{=}{\left(t^{{{2}}}\right)}^{{{\frac{1}{2\left(\mathrm{1+}\mathrm{\mu }\right)}}}}=\sum^{\mathrm{\infty }}_{k\mathrm{=0}}{\frac{{{\frac{1}{2\left(\mathrm{1+}\mathrm{\mu }\right)}}}}{{{\frac{1}{2\left(\mathrm{1+}\mathrm{\mu }\right)}}}\mathrm{+}{{\frac{1}{\mathrm{1+}\mathrm{\mu }}}}k}\left(\genfrac{}{}{0pt}{}{{{\frac{1}{2\left(\mathrm{1+}\mathrm{\mu }\right)}}}\mathrm{+}{{\frac{1}{\mathrm{1+}\mathrm{\mu }}}}k}{k}\right)}{\left(\mathrm{-}W^{{{\frac{\mathrm{2}}{\mathrm{1+}\mathrm{\mu }}}}}\right)}^k\  
\end{equation} 
and
\begin{equation} \label{GrindEQ__33_} 
t={W\left(\frac{x}{R}\right)}^{1+\mu }{=\left(t^{{{2}}}\right)}^{{{\frac{1}{\mathrm{2}}}}}\mathrm{=}\sum^{\mathrm{\infty }}_{k\mathrm{=0}}{\frac{{{\frac{1}{2}}}}{{{\frac{1}{2}}}\mathrm{+}{{\frac{1}{\mathrm{1+}\mathrm{\mu }}}}k}\left(\genfrac{}{}{0pt}{}{{{\frac{1}{2}}}\mathrm{+}{{\frac{1}{\mathrm{1+}\mathrm{\mu }}}}k}{k}\right)}{\left(\mathrm{-}W^{{{\frac{\mathrm{2}}{\mathrm{1+}\mathrm{\mu }}}}}\right)}^k\  .  
\end{equation} 

As (see (22)) \textit{D}=\textit{ W/}((1+\textit{$\mu$})${}^{1\mathrm{/}2}$ \textit{R${}^{\textrm{µ}}$}), we can again use Eq. \eqref{GrindEQ__3_} together with \eqref{GrindEQ__33_} for the calculation of the coordinate \textit{z} for the large-\textit{$\nu$} region:
\begin{equation} \label{GrindEQ__34_} 
z={Dx}^{\mathrm{1+}\mathrm{\mu }}=\frac{W}{{\left(1+\mu \right)}^{1/2}R^{\mu }}\frac{R^{1+\mu }t}{W}=\frac{R}{{\left(1+\mu \right)}^{1/2}}\ \sum^{\mathrm{\infty }}_{k\mathrm{=0}}{\frac{{{\frac{1}{2}}}}{{{\frac{1}{2}}}\mathrm{+}{{\frac{1}{\mathrm{1+}\mathrm{\mu }}}}k}\left(\genfrac{}{}{0pt}{}{{{\frac{1}{2}}}\mathrm{+}{{\frac{1}{\mathrm{1+}\mathrm{\mu }}}}k}{k}\right)}{\left(\mathrm{-}W^{{{\frac{\mathrm{2}}{\mathrm{1+}\mathrm{\mu }}}}}\right)}^k .       
\end{equation} 
The coordinate \textit{x} for the large-\textit{$\nu$} region can be retrieved from \eqref{GrindEQ__32_}:
\begin{equation} \label{GrindEQ__35_} 
x=W^{{{\frac{\mathrm{1}}{\mathrm{1+}\mathrm{\mu }}}}}\ R\mathrm{\ }t^{{{\frac{1}{\mathrm{1+}\mathrm{\mu }}}}}=W^{{{\frac{\mathrm{1}}{\mathrm{1+}\mathrm{\mu }}}}}\ R\sum^{\mathrm{\infty }}_{k\mathrm{=0}}{\frac{{{\frac{1}{2\left(\mathrm{1+}\mathrm{\mu }\right)}}}}{{{\frac{1}{2\left(\mathrm{1+}\mathrm{\mu }\right)}}}\mathrm{+}{{\frac{1}{\mathrm{1+}\mathrm{\mu }}}}k}\left(\genfrac{}{}{0pt}{}{{{\frac{1}{2\left(\mathrm{1+}\mathrm{\mu }\right)}}}\mathrm{+}{{\frac{1}{\mathrm{1+}\mathrm{\mu }}}}k}{k}\right)}{\left(\mathrm{-}W^{{{\frac{\mathrm{2}}{\mathrm{1+}\mathrm{\mu }}}}}\right)}^k .        
\end{equation} 

A convergence of the resulting power series has to be carefully considered also in this case. After the substitution 1/(1+\textit{$\mu$}) for \textit{b} and  ${\mathrm{-}W}^{{{\frac{\mathrm{2}}{\mathrm{1+}\mathrm{\mu }}}}}\mathrm{\ \ \ }$for \textit{u} into the convergence limit \eqref{GrindEQ__12_}, we arrive at

\begin{equation} \label{GrindEQ__36_} 
	\begin{aligned} 
                &\left|{\mathrm{-}\left(W^{\mathrm{2}}\right)}^{{{\frac{\mathrm{1}}{\mathrm{1+}\mathrm{\mu }}}}}\right|=\frac{1}{{\left(W^{\mathrm{2}}\right)}^{{{\frac{\mathrm{1}}{\mathrm{1+}\mathrm{\mu }}}}}}<\left|\frac{{\left(\frac{1}{\mathrm{1+}\mu }\ -1\right)}^{\frac{1}{\mathrm{1+}\mu }\ -1}}{{\left(\frac{1}{\mathrm{1+}\mu }\right)}^{\frac{1}{\mathrm{1+}\mu }}}\right|\mathrm{=}\left|\frac{{\left(\frac{1}{\mathrm{1+}\mu }\ -1\right)}^{\frac{1}{\mathrm{1+}\mu }\ }}{{\left(\frac{1}{\mathrm{1+}\mu }\right)}^{\frac{1}{\mathrm{1+}\mu }}}\frac{\mathrm{1}}{\left(\frac{1}{\mathrm{1+}\mu }\ -1\right)}\right|\mathrm{=}\left|\frac{{\left(-\mu \right)}^{\frac{1}{\mathrm{1+}\mu }\ }}{\frac{-\mu }{\mathrm{1+}\mu }}\right|\mathrm{\ \ }   \\
                &{{\stackrel{}{\Rightarrow}}}\mathrm{\ \ \ \ \ \ \ \ \ }W^{\mathrm{2}}>\left|\frac{{\left(\frac{-\mu }{\mathrm{1+}\mu }\right)}^{\mathrm{1+}\mu }}{-\mu }\right|\mathrm{=}\left|\frac{{\left(-\mu \right)}^{\mu }}{{\left(\mathrm{1+}\mu \right)}^{\mathrm{1+}\mu }}\right|\mathrm{\ \ \ \ }{{\stackrel{}{\Rightarrow}}}\mathrm{\ \ \ }\left|W\right|>\sqrt{\left|\frac{{{\left(i^2\right)}^{\mu }\ \mu }^{\mu }}{{\left(\mathrm{1+}\mu \right)}^{\mathrm{1+}\mu }}\right|}\mathrm{=}\sqrt{\frac{{\mu }^{\mu }}{{\left(\mathrm{1+}\mu \right)}^{\mathrm{1+}\mu }}}\mathrm{\ \ \ \ \ \ \ } .  
	\end{aligned} 
\end{equation}

It can be seen that the solution \eqref{GrindEQ__34_}, \eqref{GrindEQ__35_} can be used for large-\textit{$\nu$} region (for \textit{$\nu$} up to \textit{$\pi$/2}), and that this large-\textit{$\nu$} region has a low-end boundary exactly fitting with the small-\textit{$\nu$} region (see Eq. (29)). The border line between the two regions is thus defined by the formula
\begin{equation} \label{GrindEQ__37_} 
\mathrm{\ \ }W_{border}\left(R,\nu \right)=\sqrt{\frac{{\mu }^{\mu }}{{\left(\mathrm{1+}\mu \right)}^{\mathrm{1+}\mu }}}  .    
\end{equation} 

\begin{figure}
	\includegraphics[width=\linewidth]{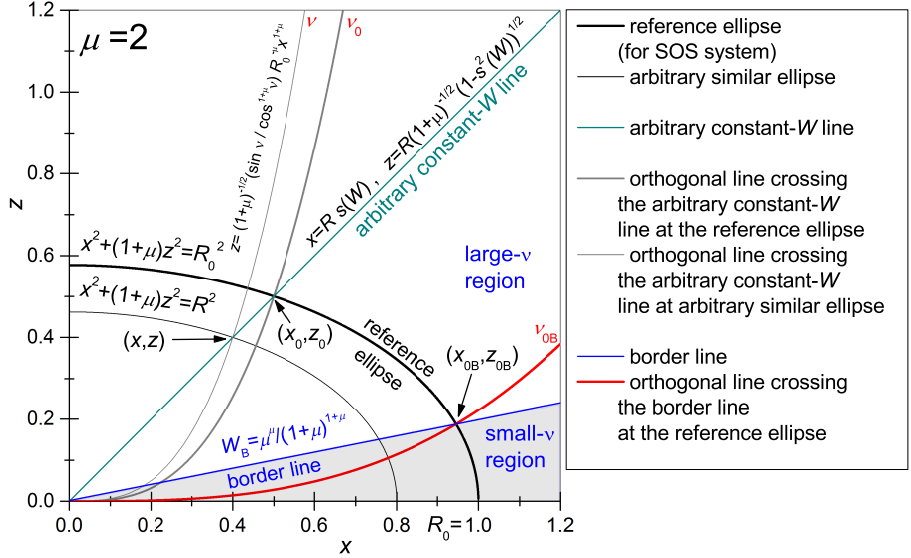}
	\caption{A One quadrant with the reference ellipse (\textit{µ}=2, \textit{R}${}_{0}$=1) of the SOS coordinate system as well as with the straight line for \textit{W}=constant radiating from the origin of coordinates in the large-$\nuup$ region. An arbitrary ellipse and two orthogonal trajectories to it having two different values of $\nuup$ coordinate are shown as well. The point (\textit{x}${}_{0}$,\textit{z}${}_{0}$) on the reference ellipse in Cartesian coordinates corresponds to the point (\textit{R}${}_{0}$,\textit{$\nu$}${}_{0}$) in SOS coordinates. Situation at and around the boundary between small- and large-$\nuup$ region is displayed as well, (\textit{x}${}_{0B}$,\textit{z}${}_{0B}$) being the point on which the boundary line and the reference ellipse intersect.}
	\label{fig:Fig2}
\end{figure}

Therefore, when computing \textit{x} and \textit{z} coordinates from the equations \eqref{GrindEQ__26_}, \eqref{GrindEQ__27_}, and \eqref{GrindEQ__34_}, \eqref{GrindEQ__35_}, a border line determined by \eqref{GrindEQ__37_} can be used for the particular point (\textit{R}, \textit{$\nu$}) in SOS coordinates as the criterion if to use the formulas \eqref{GrindEQ__26_} and \eqref{GrindEQ__27_} valid for the small-\textit{$\nu$} region, or the formulas \eqref{GrindEQ__34_} and \eqref{GrindEQ__35_} which hold for the large-\textit{$\nu$} region. \textbf{Figure \ref{fig:Fig2}} displays one quadrant with the reference ellipse (\textit{µ}=2, \textit{R}${}_{0}$=1) of the SOS coordinate system as well as other important lines of this system which will be used in further explanations, including the straight border line dividing the quadrant to the small-\textit{$\nu$} region (light gray area) and the large-\textit{$\nu$} region (white area).

It could be problematic if \textit{R} and \textit{$\nu$} coordinates are of such values that the corresponding point in \textit{x-z} plane lies exactly at the border line, as the convergence limits set by \eqref{GrindEQ__29_} and \eqref{GrindEQ__36_} seem to exclude this line. As stated earlier, \citet{Gou1956} gave the convergence limits excluding the border line. However, no detailed description of the approach leading to obtaining this result was reported in \citet{Gou1956}. In that article, the discussion of the convergence of \eqref{GrindEQ__11_} series is referenced as to be done in \citet{Pol1925}, which is, nevertheless, not the case. It seems that the Stirling`s approximation of binomial coefficients for large \textit{k} was used in \citet{Gou1956}, followed by d'Alambert's criterion for convergence. This approach was reproduced by the author. However, such approach is inconclusive on the edge of the region. \citet{Hes1957} discussed the convergence of the series \eqref{GrindEQ__11_}, particularly the case when \textit{b}$\mathrm{>}$1, \textit{y}$\rightarrow$\textit{b}/(\textit{b}$\mathrm{-}$1) and \textit{u}$\rightarrow$(\textit{b}$\mathrm{-}$1)${}^{(}$\textit{${}^{b}$}${}^{\mathrm{-}1)}$/\textit{b${}^{b}$} (in Heselden's notation: v$\mathrm{>}$1, $\mathrm{\alpha}$$\rightarrow$1/v and \textit{x}$\rightarrow$\textit{r}(v)). \citet{Hes1957} concluded that there would be a singularity for \textit{y}$\rightarrow$\textit{b}/(\textit{b}$\mathrm{-}$1) on the left side of \eqref{GrindEQ__11_} and the region of convergence thus does not include \textit{u=}(\textit{b}$\mathrm{-}$1)${}^{(}$\textit{${}^{b}$}${}^{\mathrm{-}1)}$/\textit{b${}^{b}$}. Nevertheless, he omitted the case when \textit{u} is negative (which is our case as \textit{u}=--\textit{W}${}^{2}$). The solution of the equation \eqref{GrindEQ__9_} has to be equal or less than one (\textit{y}$\leq$1). In such a case, there is no singularity when approaching {\textbar}\textit{u}{\textbar}$\rightarrow$(\textit{b}$\mathrm{-}$1)${}^{(}$\textit{${}^{b}$}${}^{\mathrm{-}1)}$/\textit{b${}^{b}$} and this argument for non-convergence of the series is thus not valid. Another source \citep{LIT2020} states that the series of \eqref{GrindEQ__10_} type converge also on the border. However, detailed derivation was neither carried out nor referenced. Therefore, the convergence/divergence issue on the border \eqref{GrindEQ__37_} between the small- and the large-\textit{$\nu$} regions is not yet concluded and it is recommended that it is further investigated by experienced mathematicians.

\noindent 
\section{ Relations for numerical transformation of \textit{R} and $\boldsymbol{\nu}$ SOS coordinates to \textit{x} and \textit{z}}

For prospective numerical calculations, it does worth to notice that there is in fact a simple dependence of \textit{x} on \textit{R}. When we look at expressions \eqref{GrindEQ__24_}, we can see that -- for a given value of the parameter \textit{W }--\textit{ }the Cartesian coordinate \textit{x} simply scales with the \textit{R} coordinate of SOS coordinate system in the small-\textit{$\nu$} region:
\begin{equation} \label{GrindEQ__38_} 
x=R\sum^{\mathrm{\infty }}_{k\mathrm{=0}}{\left(\genfrac{}{}{0pt}{}{{{\frac{1}{2}}}\mathrm{+}\left(1+\mu \right)k}{k}\right)\frac{\frac{1}{2}}{{{\frac{1}{2}}}\mathrm{+}\left(1+\mu \right)k}}{\left(\mathrm{-}W^{\mathrm{2}}\right)}^k=R\ s(W)\  .   
\end{equation} 
where \textit{s}(\textit{W}) is a parameter specific for each value of \textit{W}. Scaling also holds for \textit{z} coordinate and the proportionality parameter can be determined simply thanks to the relation \eqref{GrindEQ__1_}:
\begin{equation} \label{GrindEQ__39_} 
z=R\sqrt{\frac{\mathrm{1-}{\left[s(W)\right]}^2}{\mathrm{1+}\mu }}\mathrm{\ \ \ \ \ \ }\ \ \  .     
\end{equation} 

Thus, for a specific value of \textit{W}, the \textit{x} and \textit{z} coordinates lie on the straight line passing through the origin of the coordinate system. Then, in order to calculate numerically \textit{x} and \textit{z} for given \textit{R} and \textit{$\nu$} coordinates, it suffices to know dependence of \textit{s} on \textit{W}. The value of \textit{W} can be calculated using \eqref{GrindEQ__22_} in the first quadrant on the reference ellipse (having the large semi-axis \textit{R}${}_{0}$) with a sufficient coverage for all positions on it (i.e. for sufficiently dense coverage of \textit{$\nu$} coordinate). \textit{s}(\textit{W}) for such list of \textit{W} can be then determined e.g. using -- for small-\textit{$\nu$} region - the sum in \eqref{GrindEQ__38_} (or in other way, see later). \textit{s}(\textit{W}) can thus be tabulated in dependence of \textit{W}. At places in \textit{x-z} plane other than on the reference ellipse, \textit{W} is then calculated using \eqref{GrindEQ__22_} for given \textit{R} and \textit{$\nu$} coordinates, and the corresponding tabulated value of \textit{s}(\textit{W}) is retrieved. Multiplication by \textit{R} (see (38) and \eqref{GrindEQ__39_}) then provides \textit{x} and \textit{z} coordinates without a need to calculate the infinite power series sum again.

Moreover, there is even easier way to determine \textit{s}(\textit{W}) than by calculating the sum in \eqref{GrindEQ__38_}. The sum in \eqref{GrindEQ__38_} is connected with the equation \eqref{GrindEQ__23_}, which can be rewritten to the form
\begin{equation} \label{GrindEQ__40_} 
\mathrm{-}W^{\mathrm{2}}=\frac{{\left[s(W)\right]}^2\mathrm{-}\mathrm{1}}{{\left[{\left[s(W)\right]}^2\right]}^{\mathrm{1+}\mu }}\mathrm{\ \ \ \ \ \ }\ \ \  .     
\end{equation} 
Using \eqref{GrindEQ__22_} and \eqref{GrindEQ__40_}, we can then write for the points at the reference ellipse (i.e. with \textit{R}${}_{0}$ major semi-axis) the equation
\begin{equation} \label{GrindEQ__41_} 
\mathrm{-}W^{\mathrm{2}}=-{\left(\frac{R_0}{\mathrm{\ }R_0}\right)}^{2\mu }\frac{{\mathrm{sin}}^{\mathrm{2}}{\mathrm{\nu }}_0}{{\left({\mathrm{cos}}^{\mathrm{2}}\mathrm{\ }{\mathrm{\nu }}_0\right)}^{\mathrm{1+}\mathrm{\mu }}}=\frac{{\mathrm{cos}}^{\mathrm{2}}{\mathrm{\nu }}_0\mathrm{\ -\ 1}}{{\left({\mathrm{cos}}^{\mathrm{2}}\mathrm{\ }{\mathrm{\nu }}_0\right)}^{\mathrm{1+}\mathrm{\mu }}}=\frac{{\left[s(W)\right]}^2\mathrm{-}\mathrm{1}}{{\left[{\left[s(W)\right]}^2\right]}^{\mathrm{1+}\mu }}\ \mathrm{\ \ }\mathrm{\ \ \ \ }\ \ \  
\end{equation} 
where \textit{$\nu$}${}_{0}$ is the SOS angular coordinate value giving at the reference ellipse the same value of \textit{W} as \textit{$\nu$} coordinate value at an arbitrary similar ellipse with large semi-axis \textit{R} on which we intend to calculate corresponding \textit{x}, \textit{z} coordinates (see the points (\textit{x},\textit{z}) and (\textit{x}${}_{0}$,\textit{z}${}_{0}$) in \textbf{Figure \ref{fig:Fig2}}, where the situation is displayed for the large-\textit{$\nu$} region; similar situation is difficult to display in \textbf{Figure \ref{fig:Fig2}} for the small-\textit{$\nu$} region due to the lack of space - nevertheless, the description is equivalent in both regions).

From \eqref{GrindEQ__41_} it follows that -- although we cannot have a closed-form explicit expression for \textit{s}(\textit{W}) in terms of \textit{W} -- it is possible to calculate both \textit{W} and \textit{s}(\textit{W}) in terms of the parameter \textit{$\nu$}${}_{0}$:
\begin{equation} \label{GrindEQ__42_} 
W=\frac{{\mathrm{sin} {\mathrm{\nu }}_0\ }}{{\left({\mathrm{cos} {\mathrm{\nu }}_0\ }\right)}^{\mathrm{1+}\mathrm{\mu }}}\ \ \ \mathrm{and}\ \ \ s\left(W\right)={\mathrm{cos} {\nu }_0\ },\ \ \ \sqrt{\frac{\mathrm{1-}{\left[s(W)\right]}^2}{\mathrm{1+}\mu }}=\frac{{\mathrm{sin} {\mathrm{\nu }}_0\ }}{\sqrt{\mathrm{1+}\mu }}\ .  
\end{equation} 
This is an expected result, as the \textit{$\nu$}${}_{0}$ coordinate is (on the reference ellipse) given by the parametric definition of the ellipse.

Thus, the dependence of \textit{s}(\textit{W}) on \textit{W} can be easily tabulated and stored, for given \textit{µ}, for fast numerical transformations of \textit{R} and \textit{$\nu$} SOS coordinates to Cartesian coordinates \textit{x} and \textit{z} using \eqref{GrindEQ__38_} and \eqref{GrindEQ__39_} expressions. The numerical procedure is the following: (\textit{i}) Calculate, using \eqref{GrindEQ__22_}, \textit{W} parameter for given \textit{R} and \textit{$\nu$}; (\textit{ii}) Calculate \textit{s}(\textit{W}) corresponding to that \textit{W} using the table mapping \textit{W} to \textit{s}(\textit{W}) provided by \eqref{GrindEQ__42_}; (\textit{iii}) use \eqref{GrindEQ__38_} and \eqref{GrindEQ__39_} for the Cartesian coordinate calculation.

In the large-\textit{$\nu$} region, similar analysis can be carried out as in the small-\textit{$\nu$} region, only algebraically slightly more complicated. \eqref{GrindEQ__32_} expression provides the relation
\begin{equation} \label{GrindEQ__43_} 
x=\mathrm{\ }W^{{{\frac{\mathrm{1}}{\mathrm{1+}\mathrm{\mu }}}}}\ R\ \sum^{\mathrm{\infty }}_{k\mathrm{=0}}{\left(\genfrac{}{}{0pt}{}{{{\frac{\mathrm{1}}{\mathrm{2}\left(1+\mu \right)}}}\mathrm{+}{{\frac{\mathrm{1}}{1+\mu }}}k}{k}\right)\frac{{{\frac{\mathrm{1}}{\mathrm{2}\left(1+\mu \right)}}}}{{{\frac{\mathrm{1}}{\mathrm{2}\left(1+\mu \right)}}}\mathrm{+}{{\frac{\mathrm{1}}{1+\mu }}}k}{\left(\mathrm{-}W^{\mathrm{-}\mathrm{\ }\frac{\mathrm{2}}{\mathrm{1+}\mathrm{\mu }}}\right)}^k\mathrm{=}R\ {W^{{{\frac{\mathrm{1}}{\mathrm{1+}\mathrm{\mu }}}}}\ \left[t(W)\right]}^{\frac{\mathrm{2}}{\mathrm{2}\left(1+\mu \right)}}}\  .   
\end{equation} 
where \textit{t}(\textit{W}) is again a parameter specific for each value of \textit{W}. Scaling also holds for \textit{z} coordinate and the proportionality parameter can be determined thanks to the relation \eqref{GrindEQ__1_}:
\begin{equation} \label{GrindEQ__44_} 
z=R\ \sqrt{\frac{\mathrm{1-}{\left[W^{{{\frac{\mathrm{1}}{\mathrm{1+}\mathrm{\mu }}}}}\ {\left[t(W)\right]}^{\frac{\mathrm{1}}{1+\mu }}\right]}^2}{\mathrm{1+}\mu }}\mathrm{\ \ }\ .    
\end{equation} 

\citet{Pol1925} combinatorial identity \eqref{GrindEQ__9_} tells that the following relation is valid:
\begin{equation} \label{GrindEQ__45_} 
\mathrm{-}W^{\mathrm{-}\mathrm{\ }\frac{\mathrm{2}}{\mathrm{1+}\mathrm{\mu }}}=\frac{{\left[t(W)\right]}^{\mathrm{2}}\mathrm{-}\mathrm{1}}{{\left[t(W)\right]}^{\frac{\mathrm{2}}{1+\mu }}}\mathrm{\ \ \ \ \ \ \ \ \ \ \ }{{\stackrel{}{\Rightarrow}}}\mathrm{\ \ \ \ \ \ \ \ \ }W^{\mathrm{2}}=\frac{\mathrm{1-}\left(\mathrm{1-}{\left[t(W)\right]}^{\mathrm{2}}\right)}{{\left(\mathrm{1-}{\left[t(W)\right]}^{\mathrm{2}}\right)}^{1+\mu }}\ .   
\end{equation} 
Thus, on the reference ellipse (i.e. when the major semi-axis\textit{ R=R}${}_{0}$), we can write
\begin{equation} \label{GrindEQ__46_} 
\mathrm{-}W^{\mathrm{2}}=-\frac{{\mathrm{1-cos}}^{\mathrm{2}}{\mathrm{\nu }}_0}{{\left({\mathrm{cos}}^{\mathrm{2}}\mathrm{\ }{\mathrm{\nu }}_0\right)}^{\mathrm{1+}\mathrm{\mu }}}=-\frac{\mathrm{1-}\left(\mathrm{1-}{\left[t(W)\right]}^{\mathrm{2}}\right)}{{\left(\mathrm{1-}{\left[t(W)\right]}^{\mathrm{2}}\right)}^{1+\mu }}\ \mathrm{\ \ \ \ \ \ }\ \ \  
\end{equation} 
and then
\begin{equation} \label{GrindEQ__47_} 
\mathrm{1-}{\left[t(W)\right]}^{\mathrm{2}}={\mathrm{cos}}^{\mathrm{2}}{\mathrm{\nu }}_0\mathrm{\ \ \ \ \ \ }{{\stackrel{}{\Rightarrow}}}\mathrm{\ \ \ \ \ } t(W)={\mathrm{sin}}{\mathrm{\nu }}_0\  .   
\end{equation} 

Finally, the proportionality factor in \eqref{GrindEQ__43_} (the relation can be used also in (44)) is

\begin{equation} \label{GrindEQ__48_} 
	\begin{aligned} 
&{W^{{{\frac{\mathrm{1}}{\mathrm{1+}\mathrm{\mu }}}}}\ \left[t(W)\right]}^{\frac{\mathrm{2}}{\mathrm{2}\left(1+\mu \right)}}\mathrm{=}{\left[\frac{\mathrm{sin}{\mathrm{\ }\mathrm{\nu }}_0}{{\left(\mathrm{cos\ }{\mathrm{\nu }}_0\right)}^{\mathrm{1+}\mathrm{\mu }}}\right]}^{\mathrm{-}\mathrm{\ }\frac{\mathrm{1}}{\mathrm{1+}\mathrm{\mu }}}{\left({\mathrm{sin}}^{\mathrm{2}}{\mathrm{\nu }}_0\right)}^{\frac{\mathrm{1}}{\mathrm{2}\left(1+\mu \right)}}\mathrm{=cos\ }{\mathrm{\nu }}_0 \ ,  \\
&\sqrt{\frac{\mathrm{1-}{\left[W^{{{\frac{\mathrm{1}}{\mathrm{1+}\mathrm{\mu }}}}}\ {\left[t(W)\right]}^{\frac{\mathrm{1}}{1+\mu }}\right]}^2}{\mathrm{1+}\mu }}=\frac{{\mathrm{sin} {\mathrm{\nu }}_0\ }}{\sqrt{\mathrm{1+}\mu }}\ .  
	\end{aligned} 
\end{equation} 

Therefore, the proportionality factors for \eqref{GrindEQ__43_} and \eqref{GrindEQ__44_}, respectivelly, are -- for the large-\textit{$\nu$} region -- of the same form as the proportionality factors for \eqref{GrindEQ__38_} and \eqref{GrindEQ__39_}, respectivelly, in the small-\textit{$\nu$} region. Then, in both regions, \eqref{GrindEQ__38_} and \eqref{GrindEQ__39_} with \textit{W} to \textit{s}(\textit{W}) tabulation by \eqref{GrindEQ__42_} is to be used for numerical determination of \textit{x} and \textit{z} from \textit{R} and \textit{$\nu$}.

The situation is graphically displayed in \textbf{Figure \ref{fig:Fig2}} which shows the reference ellipse as well as the straight line for \textit{W}=\textit{constant} radiating from the origin of coordinates in the large-\textit{$\nu$} region, as well as an arbitrary ellipse and two orthogonal trajectories to it having two different values of \textit{$\nu$} coordinate. In \textbf{Figure \ref{fig:Fig2}}, the point (\textit{x}${}_{0}$,\textit{z}${}_{0}$) on the reference ellipse in Cartesian coordinates corresponds to the point (\textit{R}${}_{0}$,\textit{$\nu$}${}_{0}$) in SOS coordinates. The displayed point (\textit{x},\textit{z}) -- (\textit{R},\textit{$\nu$}) in SOS coordinates -- on the arbitrary ellipse similar to the reference one has the same value of \textit{W} parameter as the point (\textit{x}${}_{0}$,\textit{z}${}_{0}$), therefore lies on the same straight line through the origin.

\noindent 
\section{ Transformation at the border line}

Although it is important to have a simple numerical way how to transform SOS coordinates to Cartesian coordinates, as shown in the previous chapter, our main aim is to have analytical solution in order to be able to derive analytically also the metric scale factors for the use in differential operators. For this purpose, the use of infinite power series is unavoidable for writing analytic relation for transformation of SOS to Cartesian coordinates. Therefore, it is of advantage to have a closer look on the transformation also exactly at the border line between small- and large-\textit{$\nu$} regions given by \eqref{GrindEQ__37_}, where the convergence of the power series is uncertain. At all other places of the first quadrant of the \textit{x}-\textit{z} plane, the power series surely converge. Relation \eqref{GrindEQ__37_} tells that the border line is defined by a specific value of \textit{W} determined solely by the parameter \textit{µ}. As we saw in the previous section, the border between the regions is then a straight line intersecting the origin. In \textbf{Figure \ref{fig:Fig2}}, it is displayed by the blue line. \textbf{Figure \ref{fig:Fig2}} also displays a special orthogonal trajectory line (its coordinate denoted \textit{$\nu$}${}_{0B}$) intersecting the border line exactly at its intersection with the reference ellipse (\textit{R=R}${}_{0}$). The point of intersection is denoted (\textit{x}${}_{0B}$, \textit{z}${}_{0B}$) in \textbf{Figure \ref{fig:Fig2}}.

Moreover, \textit{s}(\textit{W}) also has a specific value for the border line, which is denoted \textit{s}${}_{B}$=\textit{ s}${}_{B}$(\textit{µ}) in what follows. This value can be calculated once for ever for the specific value of \textit{µ} (i.e. for the particular SOS coordinate system) using \eqref{GrindEQ__42_} and \eqref{GrindEQ__37_}. First, \textit{$\nu$}${}_{0B}$ (the anglular coordinate for the point where the reference ellipse crosses the border line) is found iteratively form the known \textit{µ} using the relation\textit{}
\begin{equation} \label{GrindEQ__49_} 
\frac{{\mu }^{\mu }}{{\left(\mathrm{1+}\mu \right)}^{\mathrm{1+}\mu }}=\frac{{{\mathrm{sin}}^{\mathrm{2}} {\mathrm{\nu }}_{\mathrm{0B}}\ }}{{\left({{\mathrm{cos}}^{\mathrm{2}} {\mathrm{\nu }}_{\mathrm{0B}}\ }\right)}^{\mathrm{1+}\mathrm{\mu }}} .     
\end{equation} 
Then, 
\begin{equation} \label{GrindEQ__50_} 
s_{0B}(\mu )={\mathrm{cos} {\mathrm{\nu }}_{\mathrm{0B}}\ }\  ,      
\end{equation} 
and, finally,
\begin{equation} \label{GrindEQ__51_} 
x=s_{0B}\left(\mu \right)R\ \ ,\ \ \  z=\sqrt{\frac{\mathrm{1-}{\left[s_{0B}\left(\mu \right)\right]}^2}{\mathrm{1+}\mu }}R\mathrm{\ \ \ }\  
\end{equation} 
on the border line.

As an example, it can be calculated that, for \textit{µ}=2 (i.e. the ellipse used in \textbf{Figure \ref{fig:Fig2}}), \textit{$\nu$}${}_{0B}$=0.331446717 rad (i.e. 18.99049802$\mathrm{{}^\circ}$) and \textit{s}${}_{B}$\eqref{GrindEQ__2_}=0.945572555. For the \textit{µ}=0.006739496742 value, corresponding to the reference spheroid of the Earth,\linebreak \textit{$\nu$}${}_{0B}$=0.77415419 rad (i.e. 44.35576776$\mathrm{{}^\circ}$) and \textit{s}${}_{B}$(0.006739496742)=0.715012606.

Then we have analytic transformation of SOS coordinates to the Cartesian ones for (\textit{i}) small-\textit{$\nu$} region (i.e \textit{W}${}^{2}$$\mathrm{<}$\textit{µ${}^{\textrm{µ}}$/}(\textit{µ}+1)\textit{${}^{\textrm{µ}}$}${}^{+1}$), given by the formulas \eqref{GrindEQ__26_}, \eqref{GrindEQ__27_}, for (\textit{ii}) large-\textit{$\nu$} region (i.e \textit{W}${}^{2}$$\mathrm{>}$\textit{µ${}^{\textrm{µ}}$/}(\textit{µ}+1)\textit{${}^{\textrm{µ}}$}${}^{+1}$), given by the formulas \eqref{GrindEQ__34_}, \eqref{GrindEQ__35_}, and also (\textit{iii}) for the border line between the two region (i.e \textit{W}${}^{2}$=\textit{µ${}^{\textrm{µ}}$/}(\textit{µ}+1)\textit{${}^{\textrm{µ}}$}${}^{+1}$), where \eqref{GrindEQ__51_} is to be used in case the power series does not converge on the border line.

It is, however, clear, that although the derivatives can be calculated from the derived relations for the small- and large-\textit{$\nu$} regions, they cannot be calculated using the relation \eqref{GrindEQ__51_} valid on the border line. Therefore, the metric scale factors (which are to be calculated using partial derivatives) has to be later again carefully assessed on the border line.

\noindent 
\section{ 3D SOS coordinates and inverse transformation}

In order to use later the known combinatorial identities for power series, it is useful to rewrite the formulas \eqref{GrindEQ__26_}, \eqref{GrindEQ__27_}, \eqref{GrindEQ__34_} and \eqref{GrindEQ__35_} also to two other equivalent forms. Important is that the shapes of the obtained expressions enable further calculation of derivatives, followed by a simplification of the resulting scale factors calculated according to the formulas \eqref{GrindEQ__16_}. This rewriting is done in the \textbf{Supplement A}, where the expressions for both \textit{x} and \textit{z} coordinate and for both series-convergence regions are calculated in two other distinct forms -- see the expressions (A5b,c), (A6b,c), (A7b,c) and (A8b,c) in the \textbf{Supplement A}. The results of these derivations are summarized in \textbf{Table 1}. These forms will be used in further derivations.  \\ \\

\noindent \textbf{Table 1}. Formulas and their parameters for Cartesian coordinates \textit{x} and \textit{z} calculation from SOS coordinates \textit{R} and \textit{$\nu$}. \textit{W} is given by \eqref{GrindEQ__22_}.

\begin{tabular}{|p{0.3in}|p{0.3in}|p{0.6in}|p{0.4in}|p{0.6in}|p{0.3in}|p{0.3in}|p{0.4in}|p{0.7in}|p{0.6in}|} \hline 
 & \textit{} &  &  & \multicolumn{4}{|p{1.6in}|}{2${}^{nd}$ type series solution} & \multicolumn{2}{|p{1.3in}|}{3${}^{rd}$ type series solution} \\ \hline 
Region & coor-dinate & scale factor for all types of series & expo-nent & \multicolumn{4}{|p{1.6in}|}{$C\frac{R}{2}\sum^{\mathrm{\infty }}_{k\mathrm{=0}}{\left(\genfrac{}{}{0pt}{}{{\alpha }_1\mathrm{+}{\mathrm{\beta }}_{\mathrm{1}}k}{k}\right)\ \frac{{\left(\mathrm{-}W^{\mathrm{\varepsilon }}\right)}^k}{{\alpha }_2\mathrm{+}{\mathrm{\beta }}_{\mathrm{2}}k}}$} & \multicolumn{2}{|p{1.3in}|}{$C\frac{R}{2}\sum^{\mathrm{\infty }}_{k\mathrm{=0}}{\left(\genfrac{}{}{0pt}{}{\alpha +\mathrm{\beta }k}{k}\right)\frac{\mathrm{-}{\left(W^{\mathrm{\varepsilon }}\right)}^k}{\alpha \mathrm{+}\mathrm{\beta }k}}$} \\ \hline 
 & \textit{} & $C$ & $\varepsilon $ & ${\alpha }_1$ & ${\mathrm{\beta }}_{\mathrm{1}}$ & ${\alpha }_2$ & ${\mathrm{\beta }}_{\mathrm{2}}$ & $\alpha $ & $\mathrm{\beta }$ \\ \hline 
small-\textit{$\nu$} region & \textit{z} & $W{\left(1+\mu \right)}^{1/2}$ & $2$ & ${{\frac{1+\mu }{2}}}\mathrm{-}\mathrm{1}$  & $1+\mu $  & $\frac{1+\mu }{2}$  & $\mu $ & $-\frac{1+\mu }{2}$  & $\mathrm{-}\mu $  \\ \hline 
 & \textit{x} & 1 & $2$ & $\mathrm{-}\frac{1}{2}$  & $1+\mu $  & $\frac{1}{2}$  & $\mu $ & $\mathrm{-}\mathrm{\ }\frac{1}{2}$  & $\mathrm{-}\mu $  \\ \hline 
large-\textit{$\nu$} region & \textit{z} & $\frac{1}{{\left(1+\mu \right)}^{1/2}}$  & $\mathrm{-}\frac{\mathrm{2}}{\mathrm{1+}\mathrm{\mu }}$  & $\mathrm{-}{{\frac{\mathrm{1}}{\mathrm{2}}}}$ & ${{\frac{1}{1+\mu }}}$ & $\frac{\mathrm{1}}{\mathrm{2}}$  & $\mathrm{-}\mathrm{\ }{{\frac{\mu }{1+\mu }}}\mathrm{\ }$ & $\mathrm{-}\mathrm{\ }\frac{1}{2}$  & ${{\frac{\mu }{1+\mu }}}$ \\ \hline 
 & \textit{x} & $\frac{W^{{{\frac{\mathrm{1}}{\mathrm{1+}\mathrm{\mu }}}}}}{1+\mu }$  & $\mathrm{-}\frac{\mathrm{2}}{\mathrm{1+}\mathrm{\mu }}$  & ${{\frac{\mathrm{1}}{\mathrm{2}\left(1+\mu \right)}}}\mathrm{-}\mathrm{1}$  & $\frac{1}{1+\mu }$  & $\frac{\mathrm{1}}{\mathrm{2}\left(1+\mu \right)}$  & $\mathrm{-}{{\frac{\mu }{1+\mu }}}\mathrm{\ }$ & $-\frac{\mathrm{1}}{\mathrm{2}\left(1+\mu \right)}$  & ${{\frac{\mu }{1+\mu }}}$ \\ \hline 
\end{tabular}  \\ \\

\noindent 
\subsection{ 3D SOS coordinates}

The 3D Cartesian coordinates \textit{x}${}_{3D}$, \textit{y}${}_{3D}$, \textit{z}${}_{3D}$ in terms of 3D similar oblate spheroidal coordinates can be obtained using \eqref{GrindEQ__15_} and Table 1 (3${}^{rd}$ type series solution) as

\begin{equation} \label{GrindEQ__52_} 
	\begin{aligned} 
&x_{\mathrm{3D}}\mathrm{=}{\mathrm{cos} \lambda \ }\ \frac{R}{2}\sum^{\mathrm{\infty }}_{k\mathrm{=0}}{\left(\genfrac{}{}{0pt}{}{-\ {{\frac{1}{2}}}\mathrm{-}\mu k}{k}\right)\frac{\mathrm{-}{\left(W^{\mathrm{2}}\right)}^k}{-\ {{\frac{1}{2}}}\mathrm{-}\mu k}\ },\ \ \ y_{\mathrm{3D}}\mathrm{=}{\mathrm{sin} \lambda \ }\ \frac{R}{2}\sum^{\mathrm{\infty }}_{k\mathrm{=0}}{\left(\genfrac{}{}{0pt}{}{-\ {{\frac{1}{2}}}\mathrm{-}\mu k}{k}\right)\frac{\mathrm{-}{\left(W^{\mathrm{2}}\right)}^k}{-\ {{\frac{1}{2}}}\mathrm{-}\mu k}}\mathrm{\ },\ \  \\
&z_{\mathrm{3D}}\mathrm{=\ }W{\left(1+\mu \right)}^{1/2}\frac{R}{2}\sum^{\mathrm{\infty }}_{k\mathrm{=0}}{\left(\genfrac{}{}{0pt}{}{-\ {{\frac{1+\mu }{2}}}\mathrm{-}\mu k}{k}\right)\frac{\mathrm{-}{\left(W^{\mathrm{2}}\right)}^k}{-\ {{\frac{1+\mu }{2}}}\mathrm{-}\mu k}} 
	\end{aligned} 
\end{equation} 
for the small-\textit{$\nu$} region, and as 

\begin{equation} \label{GrindEQ__53_} 
	\begin{aligned} 
&x_{\mathrm{3D}}\mathrm{=}{\mathrm{cos} \lambda \ }\ \frac{W^{\mathrm{-\ }{{\frac{\mathrm{1}}{\mathrm{1+}\mathrm{\mu }}}}}}{1+\mu }\ \frac{R}{2}\sum^{\mathrm{\infty }}_{k\mathrm{=0}}{\left(\genfrac{}{}{0pt}{}{-\ {{\frac{\mathrm{1}}{\mathrm{2}\left(1+\mu \right)}}}\mathrm{+}{{\frac{\mu }{1+\mu }}}k}{k}\right)\frac{-{\left(W^{\mathrm{-\ }\frac{\mathrm{2}}{\mathrm{1}\mathrm{+}\mathrm{\mu }}}\right)}^k}{-\ {{\frac{\mathrm{1}}{\mathrm{2}\left(1+\mu \right)}}}\mathrm{+}{{\frac{\mu }{1+\mu }}}k}\ },\ \  \\ 
&y_{\mathrm{3D}}\mathrm{=}{\mathrm{sin} \lambda \ }\ \frac{W^{\mathrm{-\ }{{\frac{\mathrm{1}}{\mathrm{1+}\mathrm{\mu }}}}}}{1+\mu }\ \frac{R}{2}\sum^{\mathrm{\infty }}_{k\mathrm{=0}}{\left(\genfrac{}{}{0pt}{}{-\ {{\frac{\mathrm{1}}{\mathrm{2}\left(1+\mu \right)}}}\mathrm{+}{{\frac{\mu }{1+\mu }}}k}{k}\right)\frac{-{\left(W^{\mathrm{-}\mathrm{\ }\frac{\mathrm{2}}{\mathrm{1+}\mathrm{\mu }}}\right)}^k}{-\ {{\frac{\mathrm{1}}{\mathrm{2}\left(1+\mu \right)}}}\mathrm{+}{{\frac{\mu }{1+\mu }}}k}\ },\ \  \\ 
&z_{\mathrm{3D}}\mathrm{=}\frac{1}{{\left(1+\mu \right)}^{1/2}}\ \frac{R}{2}\sum^{\mathrm{\infty }}_{k\mathrm{=0}}{\left(\genfrac{}{}{0pt}{}{{{\mathrm{-\ }\frac{\mathrm{1}}{\mathrm{2}}}}\mathrm{+}{{\frac{\mathrm{\mu}}{1+\mu }}}k}{k}\right)\frac{-{\left(W^{\mathrm{-}\mathrm{\ }\frac{\mathrm{2}}{\mathrm{1+}\mathrm{\mu }}}\right)}^k}{{{\mathrm{-\ }\frac{\mathrm{1}}{\mathrm{2}}}}\mathrm{+}{{\frac{\mathrm{\mu}}{1+\mu }}}k}}\  
	\end{aligned} 
\end{equation} 
for the large-\textit{$\nu$} region. The parameter \textit{W} is given by \eqref{GrindEQ__22_}, which shows its form for (\textit{R}, \textit{$\nu$}, \textit{$\lambda$}), or \eqref{GrindEQ__28_} for (\textit{R}, $\mathrm{\phi}$, \textit{$\lambda$}) options of the SOS coordinates. A test of the correctness of the derivation of the coordinate transformation can be done with a help of the basic formula \eqref{GrindEQ__1_} for the ellipsoid: $x^2_{3\mathrm{D}}+y^2_{3\mathrm{D}}\mathrm{+}\left(\mathrm{1+}\mathrm{\mu }\right)z^2_{3\mathrm{D}}\ $ expression determined using the above series has to be equal to \textit{R}${}^{2}$. This test was performed with the help of Cauchy product, Hagen-Rothe identity \citep{Chu2010} and the binomial identity (A4), and the result is indeed \textit{R}${}^{2}$.

\noindent 

\noindent 
\subsection{ Inverse transformation}

\noindent On the other hand, the (\textit{R}, $\mathrm{\varphi}$, \textit{$\lambda$}) or (\textit{R}, \textit{$\nu$}, \textit{$\lambda$}) coordinates can be calculated from the Cartesian coordinates as follows. From \eqref{GrindEQ__15_}, we calculate
\begin{equation} \label{GrindEQ__54_} 
\lambda ={\mathrm{arctan} \frac{y_{3\mathrm{D}}}{x_{3\mathrm{D}}}\ } .      
\end{equation} 
From \eqref{GrindEQ__1_}, we then conclude that 
\begin{equation} \label{GrindEQ__55_} 
{R= \sqrt{x^2_{3\mathrm{D}}\mathrm{+}y^2_{3\mathrm{D}}\mathrm{+}\left(\mathrm{1+}\mathrm{\mu }\right)z^2_{3\mathrm{D}}}\ }\ \ ,     
\end{equation} 
and from \eqref{GrindEQ__3_} and \eqref{GrindEQ__4_}, we have 
\begin{equation} \label{GrindEQ__56_} 
\mathrm{\varphi }\mathrm{=}{\mathrm{arctan} \frac{{\left[\sqrt{\mathrm{1+}\mathrm{\mu }\mathrm{\ \ }}a^{\mu }_0z_{3\mathrm{D}}\right]}^{\frac{\mathrm{1}}{\mathrm{1+}\mathrm{\mu }}}}{\sqrt{x^2_{3\mathrm{D}}\mathrm{+}y^2_{3\mathrm{D}}}}\ } .    
\end{equation} 

In case we use \textit{$\nu$} coordinate instead of $\mathrm{\varphi}$, the derivation of the inverse transformation formula is more complicated. Using \eqref{GrindEQ__3_} and \eqref{GrindEQ__5_}, we arrive at 
\begin{equation} \label{GrindEQ__57_} 
\frac{{\mathrm{\ \ \ sin}}\mathrm{\nu }}{{\mathrm{\ \ \ cos}}^{\mathrm{1+}\mathrm{\mu }}\mathrm{\ }\mathrm{\nu }}\mathrm{=}{\left(\mathrm{1+}\mathrm{\mu }\right)}^{{\mathrm{1}}/{\mathrm{2}}}\mathrm{\ }R^{\mu }_0\ {\frac{z_{3\mathrm{D}}}{{\left(\sqrt{x^2_{3\mathrm{D}}\mathrm{+}y^2_{3\mathrm{D}}}\right)}^{\mathrm{1+}\mathrm{\mu }}} \equiv \sqrt{A}\ }\mathrm{\ \ \ \ }\  
\end{equation} 
and then by squaring it at
\begin{equation} \label{GrindEQ__58_} 
\mathrm{\ }\ {{\frac{{\mathrm{\ \ \ sin}}^{\mathrm{2}}\mathrm{\nu }}{{\left({\mathrm{cos}}^{\mathrm{2}}\mathrm{\nu }\right)}^{\mathrm{1+}\mathrm{\mu }}}=\frac{\mathrm{1-\ }{\mathrm{cos}}^{\mathrm{2}}\mathrm{\nu }}{{\left({\mathrm{cos}}^{\mathrm{2}}\mathrm{\nu }\right)}^{\mathrm{1+}\mathrm{\mu }}}=}}\ A\ \ \ \ \ \ \ {{\stackrel{}{\Rightarrow}}}\ \ \ \ \ \ A{\left({\mathrm{cos}}^{\mathrm{2}}\mathrm{\nu }\right)}^{\mathrm{1+}\mathrm{\mu }}\mathrm{+}{\mathrm{cos}}^{\mathrm{2}}\mathrm{\nu }\mathrm{=1} .  
\end{equation} 
With a help of the substitution ${{s_A={\mathrm{cos} \mathrm{\nu }\ }\mathrm{\ }}}$, the equation is transformed to the form similar to \eqref{GrindEQ__23_},
\begin{equation} \label{GrindEQ__59_} 
{A\ s^{\mathrm{2+2}\mathrm{\mu }}_A\mathrm{+}s^{\mathrm{2}}_A =\ }\ 1\  ,      
\end{equation} 
which resembles the equation \eqref{GrindEQ__9_} and -- according to \eqref{GrindEQ__10_} -- the solution for \textit{$\nu$} in the form
\begin{equation} \label{GrindEQ__60_} 
\mathrm{\nu }=\mathrm{ar}\mathrm{cos}\left[\frac{1}{2}\sum^{\mathrm{\infty }}_{k\mathrm{=0}}{\left(\genfrac{}{}{0pt}{}{-\ {{\frac{1}{2}}}\mathrm{-}\mu k}{k}\right)\frac{\mathrm{-}A^k}{-\ {{\frac{1}{2}}}\mathrm{-}\mu k}}\right]\mathrm{\ \ \ \ \ where\ \ \ \ }A=\left(\mathrm{1+}\mathrm{\mu }\right)\mathrm{\ }R^{2\mu }_0\frac{z^2_{3\mathrm{D}}}{{\left(x^2_{3\mathrm{D}}\mathrm{+}y^2_{3\mathrm{D}}\right)}^{\mathrm{1+}\mathrm{\mu }}} 
\end{equation} 
can be thus written. The convergence of the above power series has to be found. We can use the convergence limit \eqref{GrindEQ__12_} for this purpose, with which we arrive at 
\begin{equation} \label{GrindEQ__61_} 
\left|A\right|<\frac{{\mu }^{\mu }}{{\left(\mathrm{1+}\mu \right)}^{\mathrm{1+}\mu }}\mathrm{=}\frac{\mathrm{1}}{\left.\mathrm{1+}\mu \right.}{\left(\frac{\mu }{\mathrm{1+}\mu }\right)}^{\mu }\mathrm{\ } ,     
\end{equation} 

The limit \eqref{GrindEQ__61_} for \textit{A} shows that there is a maximum of the parameter \textit{A} for which the formula \eqref{GrindEQ__60_} can be employed. It can be seen that, e.g., for large ratios ${z^2_{3\mathrm{D}}}/{{\left(x^2_{3\mathrm{D}}\mathrm{+}y^2_{3\mathrm{D}}\right)}^{\mathrm{1+}\mathrm{\mu }}}$ the series cannot converge. Therefore, for larger values of \textit{A}, another solution of \eqref{GrindEQ__59_} has to be searched for. For covering also the complementary region, the equation \eqref{GrindEQ__59_} can be -- with a help of substitution $\ t^2{= A^{{{\frac{\mathrm{1}}{\mathrm{1+}\mathrm{\mu }}}}}\mathrm{\ }s^{\mathrm{2}}\ }=A^{{{\frac{\mathrm{1}}{\mathrm{1+}\mathrm{\mu }}}}}\mathrm{\ }{\mathrm{cos}}^{\mathrm{2}}\mathrm{\nu }$   -- rewritten to the form
\begin{equation} \label{GrindEQ__62_} 
{{A^{{{\frac{\mathrm{1}}{\mathrm{1+}\mathrm{\mu }}}}}\mathrm{\ }t}^{\mathrm{2}} +\ t^{\mathrm{2+2}\mu }=\ }\ 1 
\end{equation} 
with its solution according to \eqref{GrindEQ__9_} and \eqref{GrindEQ__10_}
\begin{equation} \label{GrindEQ__63_} 
\mathrm{\nu }=\mathrm{ar}\mathrm{cos}\left[\frac{A^{{{\frac{\mathrm{1}}{2\left(1+\mu \right)}}}}}{2\left(1+\mu \right)}\sum^{\mathrm{\infty }}_{k\mathrm{=0}}{\left(\genfrac{}{}{0pt}{}{-{{\frac{\mathrm{1}}{\mathrm{2}\left(1+\mu \right)}}}\mathrm{+}{{\frac{\mu }{1+\mu }}}k}{k}\right)\ \frac{\mathrm{-}{\left(A^{\mathrm{-}\mathrm{\ }\frac{\mathrm{1}}{\mathrm{1+}\mathrm{\mu }}}\right)}^k}{-{{\frac{\mathrm{1}}{\mathrm{2}\left(1+\mu \right)}}}\mathrm{+}{{\frac{\mu }{1+\mu }}}k}}\right] .   
\end{equation} 
The convergence of the resulting power series has to be carefully considered also in this case. The convergence limit \eqref{GrindEQ__12_} is again used and we arrive at 

\begin{equation} \label{GrindEQ__64_} 
	\begin{aligned} 
&\left|{-A}^{{{\frac{\mathrm{1}}{\mathrm{1+}\mathrm{\mu }}}}}\right|=\frac{1}{A^{{{\frac{\mathrm{1}}{\mathrm{1+}\mathrm{\mu }}}}}}<\left|\frac{{\left(\frac{1}{\mathrm{1+}\mu }\ -1\right)}^{\frac{1}{\mathrm{1+}\mu }\ -1}}{{\left(\frac{1}{\mathrm{1+}\mu }\right)}^{\frac{1}{\mathrm{1+}\mu }}}\right|\mathrm{=}\left|\frac{{\left(\frac{1}{\mathrm{1+}\mu }\ -1\right)}^{\frac{1}{\mathrm{1+}\mu }\ }}{{\left(\frac{1}{\mathrm{1+}\mu }\right)}^{\frac{1}{\mathrm{1+}\mu }}}\frac{\mathrm{1}}{\left(\frac{1}{\mathrm{1+}\mu }\ -1\right)}\right|\mathrm{\ } \\
&{{\stackrel{}{\Rightarrow}}}\mathrm{\ }A>\left|\frac{{{\left(i^2\right)}^{\mu }\ \mu }^{\mu }}{{\left(\mathrm{1+}\mu \right)}^{\mathrm{1+}\mu }}\right|\mathrm{=}\frac{{\mu }^{\mu }}{{\left(\mathrm{1+}\mu \right)}^{\mathrm{1+}\mu }}\mathrm{\ } .  
	\end{aligned} 
\end{equation} 

It can be seen that the complementary region has a boundary exactly fitting with the first region (Eq. (61)). The border for the use of either solution \eqref{GrindEQ__60_} or the solution \eqref{GrindEQ__63_} is thus defined by the formula
\begin{equation} \label{GrindEQ__65_} 
\mathrm{\ \ }A_{border}\left(x_{3\mathrm{D}}{,y}_{3\mathrm{D}}{,z}_{3\mathrm{D}}\right)=\frac{{\mu }^{\mu }}{{\left(\mathrm{1+}\mu \right)}^{\mathrm{1+}\mu }} .     
\end{equation}

\noindent 
\section{ Partial derivatives }

In order to calculate the metric scale factors according to \eqref{GrindEQ__16_}, we need to determine partial derivatives of \textit{x} and \textit{z} with respect to \textit{R} and \textit{$\nu$}.

\noindent 
\subsection{ Partial derivatives with respect to \textit{R}}

First, partial derivatives with respect to \textit{R} are determined. When Eq. (A6b) (i.e. the 2${}^{nd}$ type series solution, see Table 1) is used for the coordinate \textit{x} in the small-\textit{$\nu$} region, and considering dependence \eqref{GrindEQ__22_} of \textit{W} on \textit{R}${}^{\textrm{µ}}$, the partial derivative has the form
\begin{equation} \label{GrindEQ__66_} 
\frac{\partial x\left(\nu ,R\right)}{\partial R}\ \mathrm{=\ }\frac{1}{2}\sum^{\mathrm{\infty }}_{k\mathrm{=0}}{\left(\genfrac{}{}{0pt}{}{{{\frac{1}{2}}}\mathrm{+}\left(1+\mu \right)k}{k}\right)\ \frac{{\mathrm{(2}\mu k+1)\left(\mathrm{-}W^{\mathrm{2}}\right)}^k}{{{\frac{1}{2}}}\mathrm{+}\mu k}}\mathrm{=}\sum^{\mathrm{\infty }}_{k\mathrm{=0}}{\left(\genfrac{}{}{0pt}{}{{{\frac{1}{\ 2}}}\mathrm{+}\left(1+\mu \right)k}{k}\right){\left(\mathrm{-}W^{\mathrm{2}}\right)}^k\ } .  
\end{equation} 
With this choice, the terms in the denominator and numerator are mutually canceled, and the binomial coefficient has a form advantageous for a later calculation of the metric factor \textit{h}${}_{R}$. Similarly, partial derivative of \textit{z} with respect to \textit{R} in the small-\textit{$\nu$} region is obtained using Eq. (A5b) 
\begin{equation} \label{GrindEQ__67_} 
\frac{\partial z\left(\nu ,R\right)}{\partial R}\ \mathrm{=\ }W{\left(1+\mu \right)}^{1/2}\sum^{\mathrm{\infty }}_{k\mathrm{=0}}{\left(\genfrac{}{}{0pt}{}{{{\frac{\mu -1}{2}}}\mathrm{+}\left(1+\mu \right)k}{k}\right){\left(\mathrm{-}W^{\mathrm{2}}\right)}^k\ } .   
\end{equation} 

When Eq. (A8b) (i.e. the 2${}^{nd}$ type series solution, see Table 1) is used for the coordinate \textit{x} in the large-\textit{$\nu$} region, and again considering dependence \eqref{GrindEQ__22_} of \textit{W} on \textit{R}${}^{\textrm{µ}}$, the partial derivative has the form
\begin{equation} \label{GrindEQ__68_} 
\frac{\partial x\left(\nu ,R\right)}{\partial R}\ \ \mathrm{=\ }\frac{W^{{{\frac{\mathrm{1}}{\mathrm{1+}\mathrm{\mu }}}}}}{1+\mu }\ \sum^{\mathrm{\infty }}_{k\mathrm{=0}}{\left(\genfrac{}{}{0pt}{}{\left({{\frac{\mathrm{1}}{\mathrm{2}\left(1+\mu \right)}}}\mathrm{-}\mathrm{1}\right)\mathrm{+}{{\frac{1}{1+\mu }}}k}{k}\right){\left(\mathrm{-}W^{\mathrm{-}\mathrm{\ }\frac{\mathrm{2}}{\mathrm{1+}\mathrm{\mu }}}\right)}^k\ } .   
\end{equation} 
Similarly, partial derivative of \textit{z} with respect to \textit{R} in the large-\textit{$\nu$} region is obtained using Eq. (A7b) 
\begin{equation} \label{GrindEQ__69_} 
\frac{\partial z\left(\nu ,R\right)\ \ }{\partial R}\ \ \mathrm{=}\frac{1}{{\left(1+\mu \right)}^{1/2}}\ \sum^{\mathrm{\infty }}_{k\mathrm{=0}}{\left(\genfrac{}{}{0pt}{}{\mathrm{-}{{\frac{\mathrm{1}}{\mathrm{2}}}}\mathrm{+}{{\frac{1}{1+\mu }}}k}{k}\right){\left(\mathrm{-}W^{\mathrm{-}\mathrm{\ }\frac{\mathrm{2}}{\mathrm{1+}\mathrm{\mu }}}\right)}^k\ }\mathrm{\ } .   
\end{equation}

\noindent 
\subsection{ Partial derivatives with respect to $\boldsymbol{\nu}$}

Further, partial derivatives with respect to $\mathrm{\nu}$ are determined. As rather lengthy derivation is needed for simplification of the given binomial expressions (including several combinatorial identities listed in \citet{Gou1972,Chu2010,Chu2013,Sei1994}), the derivation for the small-$\mathrm{\nu}$ region is carried out in Supplement B, and for the large-$\mathrm{\nu}$ region in Supplement C.

\noindent 
\paragraph{1) Small-$\nu$  region }

Here, the final result (see \textbf{Supplement B} for details) of derivation of the expressions for the derivatives

\begin{equation} \label{GrindEQ__70_} 
	\begin{aligned} 
&\frac{\partial x\left(\nu ,R\right)}{\partial \nu }=\frac{R}{2}\ \frac{1}{\mu W}\frac{\partial W}{\partial \nu }\left\{\sum^{\mathrm{\infty }}_{k\mathrm{=0}}{\left(\genfrac{}{}{0pt}{}{-\ {{\frac{1}{2}}}\mathrm{-}\mu k}{k}\right)\ \frac{{\left(W^{\mathrm{2}}\right)}^k}{-\ {{\frac{1}{2}}}\mathrm{-}\mu k}}\mathrm{+}2\sum^{\mathrm{\infty }}_{k\mathrm{=0}}{\left(\genfrac{}{}{0pt}{}{-\ {{\frac{1}{2}}}\mathrm{-}\mu k}{k}\right){\left(W^{{{}}}\right)}^k\ }\right\}  \\ 
&=R\ \frac{1}{W}\frac{\partial W}{\partial \nu }\sum^{\mathrm{\infty }}_{k\mathrm{=1}}{\left(\genfrac{}{}{0pt}{}{-\ {{\frac{1}{2}}}\mathrm{-}\mu k}{k}\right)\ \frac{2k}{2\mu k+1}}{\left(W^{\mathrm{2}}\right)}^k\ ,   
	\end{aligned} 
\end{equation} 

\begin{equation} \label{GrindEQ__71_} 
	\begin{aligned} 
&\frac{\partial z\left(\nu ,R\right)}{\partial \nu }\mathrm{=}{\left(1+\mu \right)}^{1/2}\frac{R}{2}\ \frac{1}{\mu }\frac{\partial W}{\partial \nu }\left\{\sum^{\mathrm{\infty }}_{k\mathrm{=0}}{\left(\genfrac{}{}{0pt}{}{-\ {{\frac{1+\mu }{2}}}\mathrm{-}\mu k}{k}\right)\frac{{\left(W^{\mathrm{2}}\right)}^k}{-\ {{\frac{1+\mu }{2}}}\mathrm{-}\mu k}}\mathrm{+}2\sum^{\mathrm{\infty }}_{k\mathrm{=0}}{\left(\genfrac{}{}{0pt}{}{-\ {{\frac{1+\mu }{2}}}\mathrm{-}\mu k}{k}\right){\left(W^{{{}}}\right)}^k\ }\right\}  \\ 
&={\left(1+\mu \right)}^{1/2}R\frac{\partial W}{\partial \nu }\sum^{\mathrm{\infty }}_{k\mathrm{=0}}{\left(\genfrac{}{}{0pt}{}{-\ {{\frac{1+\mu }{2}}}\mathrm{-}\mu k}{k}\right)\frac{2k+1}{\left(2k+1\right)\mu +1}}{\left(W^{\mathrm{2}}\right)}^k\ ,  
	\end{aligned} 
\end{equation} 

\noindent and of the squared derivatives (in the form suitable for the subsequent metric scale factor calculation)

\begin{equation} \label{GrindEQ__72_} 
	\begin{aligned} 
&{\left(\frac{\partial x\left(\nu ,R\right)}{\partial \nu }\right)}^2\mathrm{=}{\left(\frac{R}{2}\right)}^2\frac{1}{{\mu }^2}{\left(\frac{\partial W}{\partial \nu }\right)}^2\left\{\mathrm{-}\mathrm{4}\frac{\mu +1}{\mu }\sum^{\mathrm{\infty }}_{k\mathrm{=1}}{\frac{1}{-\mu k-1}\left(\genfrac{}{}{0pt}{}{\mathrm{-}\mu k}{k}\right){\left(W^{\mathrm{2}}\right)}^{k-1}}\mathrm{+}\frac{\mathrm{4}}{W^{\mathrm{2}}} \right. \\ 
& \left. \mathrm{-}\mathrm{8}\frac{\mu +1}{\mu }\sum^{\mathrm{\infty }}_{k\mathrm{=1}}{\left(\genfrac{}{}{0pt}{}{\mathrm{-}\mu k}{k}\right){\left(W^{\mathrm{2}}\right)}^{k-1}}\mathrm{-}\frac{\mathrm{8}}{W^{\mathrm{2}}}\mathrm{-}\mathrm{4}\sum^{\mathrm{\infty }}_{k\mathrm{=0}}{\left[{\left({-W}^{\mathrm{2}}\right)}^{k-1}\sum^k_{m\mathrm{=0}}{\left(\genfrac{}{}{0pt}{}{\mu k+m-1}{m}\right){\left(1+\mu \right)}^{k-m}}\right]}\right\}\ ,  
	\end{aligned} 
\end{equation} 

\begin{equation} \label{GrindEQ__73_} 
	\begin{aligned} 
&{\left(\frac{\partial z\left(\nu ,R\right)}{\partial \nu }\right)}^2\mathrm{=}{\left(\frac{R}{2}\right)}^2\frac{1}{{\mu }^2}{\left(\frac{\partial W}{\partial \nu }\right)}^2\left\{\frac{\mathrm{4}}{\mu }\sum^{\mathrm{\infty }}_{M\mathrm{=1}}{\frac{1}{-\mu M-1}\left(\genfrac{}{}{0pt}{}{\mathrm{-}\mu M}{M}\right){\left(W^{\mathrm{2}}\right)}^{M-1}}\mathrm{+}\frac{\mathrm{8}}{\mu }\sum^{\mathrm{\infty }}_{M\mathrm{=1}}{\left(\genfrac{}{}{0pt}{}{\mathrm{-}\mu M}{M}\right){\left(W^{\mathrm{2}}\right)}^{M-1}} \right. \\ 
& \left. \mathrm{+4}\sum^{\mathrm{\infty }}_{M\mathrm{=0}}{\left[{\left(\mathrm{-}W^{\mathrm{2}}\right)}^{M-1}\sum^M_{m\mathrm{=0}}{\left(\genfrac{}{}{0pt}{}{\mu M+m-1}{m}\right){\left(1+\mu \right)}^{M-m}}\right]}\mathrm{-}\mathrm{4}\sum^{\mathrm{\infty }}_{M\mathrm{=0}}{{\left(\mathrm{-}W^{\mathrm{2}}\right)}^{M-1}\left(\genfrac{}{}{0pt}{}{\mu M+M-1}{M}\right)}\right\}\ ,  
	\end{aligned} 
\end{equation} 

\noindent are reported for the small-$\mathrm{\nu}$ region. It can be observed, that the squared derivatives of \textit{x} and \textit{z} contain similar terms, which can facilitate a simplification of the metric scale factor calculated later.

\textit{W} (see Eq. (22)) and its derivative appearing in the above formulas are given by
\begin{equation} \label{GrindEQ__74_} 
W={\left(\frac{R}{\mathrm{\ }R_0}\right)}^{\mu }\frac{{\mathrm{sin}}\mathrm{\nu }}{{\mathrm{\ cos}}^{\mathrm{1+}\mathrm{\mu }}\mathrm{\ }\mathrm{\nu }}\ \ \ \ \mathrm{and}\ \ \ \ \ \frac{\partial W}{\partial \nu }={\left(\frac{R}{\mathrm{\ }R_0}\right)}^{\mu }\frac{\mathrm{1+}\mathrm{\mu }\mathrm{\ }{\mathrm{sin}}^{\mathrm{2}}\mathrm{\nu }}{{\mathrm{cos}}^{\mathrm{2+}\mathrm{\mu }}\mathrm{\ }\mathrm{\nu }}\ \ \ \ \ \mathrm{and}\ \ \ \ \ \ \frac{1}{W}\frac{\partial W}{\partial \nu }=\frac{\mathrm{1+}\mathrm{\mu }\mathrm{\ }{\mathrm{sin}}^{\mathrm{2}}\mathrm{\nu }}{\mathrm{\ }{{\mathrm{sin} \mathrm{\nu }\mathrm{\ }\ }\mathrm{cos\ } \mathrm{\nu }\ }}\  .  
\end{equation} 
The derivative of \textit{W} can be derived also with respect to the \citet{Whi2008} coordinate $\mathrm{\phi}$. This can be done using \eqref{GrindEQ__28_}:
\begin{equation} \label{GrindEQ__75_} 
W=\frac{R^{\textrm{µ}}}{a^{\mu }_0}{\mathrm{tan}}^{\mathrm{1+}\mathrm{\mu }}\mathrm{\varphi }\ ,\ \ \ \frac{\partial W}{\partial \mathrm{\varphi }}=\left(\mathrm{1+}\mathrm{\mu }\right)\frac{R^{\textrm{µ}}}{a^{\mu }_0}{\mathrm{tan}}^{\mathrm{\mu }}\mathrm{\varphi }\left(\mathrm{1+}{\mathrm{tan}}^{\mathrm{2}}\mathrm{\varphi }\right)\ ,\ \ \frac{1}{W}\frac{\partial W}{\partial \mathrm{\varphi }}=\left(\mathrm{1+}\mathrm{\mu }\right)\frac{\mathrm{1+}{\mathrm{tan}}^{\mathrm{2}}\mathrm{\varphi }}{\mathrm{\ tan}\mathrm{\varphi }}\ .     
\end{equation} 

\paragraph{2) Large-$\nu$  region }

\noindent Here, the final result (see \textbf{Supplement C} for details) of derivation of the expressions for the derivatives

\begin{equation} \label{GrindEQ__76_} 
	\begin{aligned} 
&\frac{\partial x\left(\nu ,R\right)}{\partial \nu }=  \\ 
&\frac{\mathrm{1}}{1+\mu }\ \frac{R}{2}\frac{1}{\mu W^{\mathrm{1+\ }\frac{\mathrm{1}}{\mathrm{1+}\mathrm{\mu }}}}\frac{\partial W}{\partial \nu }\left\{\sum^{\mathrm{\infty }}_{k\mathrm{=0}}{\left(\genfrac{}{}{0pt}{}{-{{\frac{\mathrm{1}}{\mathrm{2}\left(1+\mu \right)}}}\mathrm{+}{{\frac{\mu }{1+\mu }}}k}{k}\right)\frac{{\left(W^{\mathrm{-}\mathrm{\ }\frac{\mathrm{2}}{\mathrm{1+}\mathrm{\mu }}}\right)}^k}{{{\frac{\mathrm{1}}{\mathrm{2}\left(1+\mu \right)}}}\mathrm{+}{{\frac{\mu }{1+\mu }}}k}}\mathrm{+}2\sum^{\mathrm{\infty }}_{k\mathrm{=0}}{\left(\genfrac{}{}{0pt}{}{-{{\frac{\mathrm{1}}{\mathrm{2}\left(1+\mu \right)}}}\mathrm{+}{{\frac{\mu }{1+\mu }}}k}{k}\right){\left(W^{\mathrm{-}\mathrm{\ }\frac{\mathrm{2}}{\mathrm{1+}\mathrm{\mu }}}\right)}^k}\right\}  \\ 
&=\frac{R}{1+\mu }\ \frac{1}{W^{\mathrm{1+\ }\frac{\mathrm{1}}{\mathrm{1+}\mathrm{\mu }}}}\frac{\partial W}{\partial \nu }\sum^{\mathrm{\infty }}_{k\mathrm{=0}}{\left(\genfrac{}{}{0pt}{}{-\ {{\frac{\mathrm{1}}{\mathrm{2}\left(1+\mu \right)}}}\mathrm{+}{{\frac{\mu }{1+\mu }}}k}{k}\right)\frac{2k+1}{2\mu k-1}}{\left(W^{\mathrm{-}\mathrm{\ }\frac{\mathrm{2}}{\mathrm{1+}\mathrm{\mu }}}\right)}^k\  ,  
	\end{aligned} 
\end{equation} 

\begin{equation} \label{GrindEQ__77_} 
	\begin{aligned} 
&\frac{\partial z\left(\nu ,R\right)}{\partial \nu }=  \\ 
&\frac{1}{{\left(1+\mu \right)}^{1/2}}\frac{R}{2}\frac{1}{\mu W}\frac{\partial W}{\partial \nu }\left\{\sum^{\mathrm{\infty }}_{k\mathrm{=0}}{\left(\genfrac{}{}{0pt}{}{-{{\frac{\mathrm{1}}{\mathrm{2}}}}\mathrm{+}{{\frac{\mu }{1+\mu }}}k}{k}\right)\ \frac{{\left(W^{\mathrm{-}\frac{\mathrm{2}}{\mathrm{1+}\mathrm{\mu }}}\right)}^k}{-{{\frac{1}{2}}}\mathrm{+}{{\frac{\mu }{1+\mu }}}k}}\mathrm{+}2\sum^{\mathrm{\infty }}_{k\mathrm{=0}}{\left(\genfrac{}{}{0pt}{}{-{{\frac{\mathrm{1}}{\mathrm{2}}}}\mathrm{+}{{\frac{\mu }{1+\mu }}}k}{k}\right){\left(W^{\mathrm{-}\frac{\mathrm{2}}{\mathrm{1+}\mathrm{\mu }}}\right)}^k}\right\}  \\ 
&=\frac{R}{{\left(1+\mu \right)}^{1/2}}\ \frac{1}{W}\frac{\partial W}{\partial \nu }\sum^{\mathrm{\infty }}_{k\mathrm{=1}}{\left(\genfrac{}{}{0pt}{}{-\ {{\frac{\mathrm{1}}{\mathrm{2}}}}\mathrm{+}{{\frac{\mu }{1+\mu }}}k}{k}\right)\ \frac{2k}{2\mu k-\mu -1\ }}{\left(W^{\mathrm{-}\mathrm{\ }\frac{\mathrm{2}}{\mathrm{1+}\mathrm{\mu }}}\right)}^k ,  
	\end{aligned} 
\end{equation} 

\noindent and of the square derivatives in the form suitable for the subsequent metric scale factor calculation

\begin{equation} \label{GrindEQ__78_} 
	\begin{aligned} 
&{\left(\frac{\partial x\left(\nu ,R\right)}{\partial \nu }\right)}^2\ \ \mathrm{=}{\left(\frac{R}{2}\right)}^2\frac{1}{{\mu }^2W^{\mathrm{2+\ }\frac{\mathrm{2}}{\mathrm{1+}\mathrm{\mu }}}}{\left(\frac{\partial W}{\partial \nu }\right)}^2\left\{\mathrm{-}\mathrm{4\ }\frac{\mathrm{1+}\mathrm{\mu }}{\mathrm{\mu }}\sum^{\mathrm{\infty }}_{M\mathrm{=1}}{\frac{1}{\mu M-\mu -1}\left(\genfrac{}{}{0pt}{}{{{\frac{\mu }{1+\mu }}}M}{M}\right){\left(W^{\mathrm{-}\mathrm{\ }\frac{\mathrm{2}}{\mathrm{1+}\mathrm{\mu }}}\right)}^{M-1}} \right. \\ 
& \left. \mathrm{-}\mathrm{\ }\frac{\mathrm{8}}{\mu }\sum^{\mathrm{\infty }}_{M\mathrm{=1}}{\left(\genfrac{}{}{0pt}{}{{{\frac{\mu }{1+\mu }}}M}{M}\right){\left(W^{\mathrm{-}\mathrm{\ }\frac{\mathrm{2}}{\mathrm{1+}\mathrm{\mu }}}\right)}^{M-1}} \right. \\ 
& \left. \mathrm{+}\frac{\mathrm{4}}{1+\mu }\sum^{\mathrm{\infty }}_{M\mathrm{=0}}{\left[{\left({-W}^{\mathrm{-}\mathrm{\ }\frac{\mathrm{2}}{\mathrm{1+}\mathrm{\mu }}}\right)}^{M-1}\left(\sum^M_{m\mathrm{=0}}{\left(\genfrac{}{}{0pt}{}{-\mathrm{\ }{{\frac{\mathrm{\mu }}{\mathrm{1+}\mathrm{\mu }}}}M-1+m}{m}\right)\frac{1}{{\left(1+\mu \right)}^{M-m}}}\mathrm{-}\left(\genfrac{}{}{0pt}{}{-\mathrm{\ }{{\frac{\mathrm{\mu }}{\mathrm{1+}\mathrm{\mu }}}}M+M-1}{M}\right)\right)\right]}\right\} ,  
	\end{aligned} 
\end{equation} 

\begin{equation} \label{GrindEQ__79_} 
	\begin{aligned} 
&{\left(\frac{\partial z\left(\nu ,R\right)}{\partial \nu }\right)}^2={\frac{1}{1+\mu }\left(\frac{R}{2}\right)}^2\frac{1}{{\mu }^2W^{\mathrm{2+\ }\frac{\mathrm{2}}{\mathrm{1+}\mathrm{\mu }}}}{\left(\frac{\partial W}{\partial \nu }\right)}^2\left\{\mathrm{\ }\frac{4}{\mu }\sum^{\mathrm{\infty }}_{k\mathrm{=1}}{\frac{1}{-1+{{\frac{\mu }{1+\mu }}}k}\left(\genfrac{}{}{0pt}{}{{{\frac{\mu }{1+\mu }}}k}{k}\right){\left(W^{\mathrm{-}\mathrm{\ }\frac{\mathrm{2}}{\mathrm{1+}\mathrm{\mu }}}\right)}^{k\mathrm{\ -1}}} \right. \\ 
& \left. \mathrm{+}\frac{\mathrm{4}}{W^{\mathrm{-}\mathrm{\ }\frac{\mathrm{2}}{\mathrm{1+}\mathrm{\mu }}}}\mathrm{\ \ \ +\ }\frac{8}{\mu }\sum^{\mathrm{\infty }}_{k\mathrm{=1}}{\left(\genfrac{}{}{0pt}{}{{{\frac{\mu }{1+\mu }}}k}{k}\right){\left(W^{\mathrm{-}\mathrm{\ }\frac{\mathrm{2}}{\mathrm{1+}\mathrm{\mu }}}\right)}^{k\mathrm{\ -1}}}\mathrm{-}\frac{\mathrm{8}}{W^{\mathrm{-}\mathrm{\ }\frac{\mathrm{2}}{\mathrm{1+}\mathrm{\mu }}}} \right. \\ 
& \left. \mathrm{\ \ -4}\sum^{\mathrm{\infty }}_{k\mathrm{=0}}{\left[{\left(-1\right)}^{k-1}{\left(W^{\mathrm{-}\mathrm{\ }\frac{\mathrm{2}}{\mathrm{1+}\mathrm{\mu }}}\right)}^{k\mathrm{\ -1}}\sum^k_{m\mathrm{=0}}{\left(\genfrac{}{}{0pt}{}{-\ {{\frac{\mu }{1+\mu }}}k-1+m}{m}\right)\frac{1}{{\left(1+\mu \right)}^{k-m}}}\right]}\right\} ,  
	\end{aligned} 
\end{equation} 

\noindent are reported for the large-$\mathrm{\nu}$ region. \textit{W} and its derivative are given by the Eq. \eqref{GrindEQ__74_}. It can be observed that the square derivatives of \textit{x} and \textit{z} contain similar terms, which can facilitate a very significant simplification of the metric scale factor formula calculated in what follows.

\noindent 
\section{ Metric scale factors}

After derivation of formulas for coordinates and for partial derivatives, all means are ready for calculation of the scale factors of SOS coordinates according to Eq. \eqref{GrindEQ__16_}. We have to calculate it separately in the small- and in the large-$\mathrm{\nu}$ regions.

\noindent 
\subsection{ \textit{h}${}_{\boldsymbol{\lambda }}$ metric scale factor}

The metric scale factor \textit{h${}_{\lambda }$}, given by Eq. \eqref{GrindEQ__16_} is -- according to (A6c) from the \textbf{Supplement A} or Table 1 -- in the small-\textit{$\nu$} region the following:
\begin{equation} \label{GrindEQ__80_} 
h_{\lambda }\ \mathrm{=\ }x\left(\nu ,R\right)\ =\ \ \frac{R}{2}\sum^{\mathrm{\infty }}_{k\mathrm{=0}}{\left(\genfrac{}{}{0pt}{}{-\ {{\frac{1}{2}}}\mathrm{-}\mu k}{k}\right)\ \frac{\mathrm{-}{\left(W^{\mathrm{2}}\right)}^k}{-\ {{\frac{1}{2}}}\mathrm{-}\mu k}\ }\ ,   
\end{equation} 
while it has, according to (A8c) or Table 1, the following form in the large-\textit{$\nu$} region:
\begin{equation} \label{GrindEQ__81_} 
\ h_{\lambda }\ \mathrm{=\ }x\left(\nu ,R\right)=\ \frac{W^{{{\frac{\mathrm{1}}{\mathrm{1+}\mathrm{\mu }}}}}}{1+\mu }\ \frac{R}{2}\sum^{\mathrm{\infty }}_{k\mathrm{=0}}{\left(\genfrac{}{}{0pt}{}{-\ {{\frac{\mathrm{1}}{\mathrm{2}\left(1+\mu \right)}}}\mathrm{+}{{\frac{\mu }{1+\mu }}}k}{k}\right)\ \frac{\mathrm{-}{\left(W^{\mathrm{-}\mathrm{\ }\frac{\mathrm{2}}{\mathrm{1+}\mathrm{\mu }}}\right)}^k}{-\ {{\frac{\mathrm{1}}{\mathrm{2}\left(1+\mu \right)}}}\mathrm{+}{{\frac{\mu }{1+\mu }}}k}}\ .   
\end{equation} 

It should be reminded that the border line between the small-\textit{$\nu$} region and the large-\textit{$\nu$} region for power series of this type is defined by the formula \eqref{GrindEQ__37_}. On the border line, \textit{h${}_{\lambda }$} can be expressed according to previously derived formulas \eqref{GrindEQ__50_} and \eqref{GrindEQ__51_} with \textit{µ}-specific constant \textit{s}${}_{B}$.

\noindent 
\subsection{ \textit{h}${}_{R}$ metric scale factor}

For the derivation of \textit{h${}_{R}$}, Eqs. \eqref{GrindEQ__66_} and \eqref{GrindEQ__67_} are used. When the derivatives are input to the basic equation \eqref{GrindEQ__16_}, we obtain -- after simplification described in detail in \textbf{Supplement D} -- the following rather simple shape for the small-\textit{$\nu$} region: 

\begin{equation} \label{GrindEQ__82_} 
h_R\mathrm{=}\ \sqrt{\mathrm{\ }\sum^{\mathrm{\infty }}_{k\mathrm{=0}}{\left(\genfrac{}{}{0pt}{}{-\mu k}{k}\right){\left(W^{\mathrm{2}}\right)}^k\ }}\ .     
\end{equation} 

Other forms of this scale factor can be obtained, e.g. with a help of (A3) or other binomial identities. These easy options are reported in \textbf{Supplement D}. Moreover, the P\'{o}lya and Szeg\"{o} identity listed in \citet{Gou1972} under the number 1.120 leads to the scale factor containing the series in the denominator rather than in the numerator. The derivation is in \textbf{Supplement D}, and the result is reported also here:
\begin{equation} \label{GrindEQ__83_} 
h_R\mathrm{=}{\left[\left(\mathrm{1+}\mu \right)+\mu \sum^{\mathrm{\infty }}_{k\mathrm{=0}}{\left(\genfrac{}{}{0pt}{}{-\mu k-1}{k}\right)\frac{{\left(W^{\mathrm{2}}\right)}^k}{-\mu k-1}\ }\right]}^{\mathrm{-}\mathrm{\ }\frac{\mathrm{1}}{\mathrm{2}}}\mathrm{\ \ }  .   
\end{equation} 

In certain cases, it could be of advantage to get rid of the square root on the right side of \eqref{GrindEQ__82_} altogether. For this purpose, the binomial theorem can be used, followed by a smart counting of infinite double-sums, which leads to a compositae-containing power series \citep{Kru2013}. Then, \textit{h${}_{R}$} can be received in a pure power-series form with coefficients\textit{ c}${}_{n}$:
\begin{equation} \label{GrindEQ__84_} 
h_R\mathrm{=}{\left(\mathrm{1+}\sum^{\mathrm{\infty }}_{k\mathrm{=1}}{\left(\genfrac{}{}{0pt}{}{-\mu k}{k}\right){\left(W^{\mathrm{2}}\right)}^k}\right)}^{\frac{\mathrm{1}}{\mathrm{2}}}=\sum^{\mathrm{\infty }}_{n\mathrm{=0}}{\left(\genfrac{}{}{0pt}{}{{{\frac{1}{2}}}}{n}\right){\left(\sum^{\mathrm{\infty }}_{k\mathrm{=1}}{\left(\genfrac{}{}{0pt}{}{-\mu k}{k}\right){\left(W^{\mathrm{2}}\right)}^k}\right)}^n}=\sum^{\mathrm{\infty }}_{n\mathrm{=0}}{c_n{\left(W^{\mathrm{2}}\right)}^n}\ .  
\end{equation} 
Details of a proper compositae determination (\citet{Kru2013}, p.13) leading to the coefficients \textit{c}${}_{n}$ are, nevertheless, out of the scope of this paper.

In the large-\textit{$\nu$} region, the simplification (again described in \textbf{Supplement D}) leads to the following form of the metric scale factor
\begin{equation} \label{GrindEQ__85_} 
{\ h}_R\mathrm{=}\ \frac{\mathrm{1}}{\sqrt{1+\mu }}\sqrt{\mathrm{\ }\sum^{\mathrm{\infty }}_{k\mathrm{=0}}{\left(\genfrac{}{}{0pt}{}{{{\frac{\mu }{\mathrm{1+}\mathrm{\mu }}}}\ k}{k}\right){\left(W^{\mathrm{-}\mathrm{\ }\frac{\mathrm{2}}{\mathrm{1+}\mathrm{\mu }}}\right)}^k}}\ \  ,    
\end{equation} 
where $W\equiv W\left(R,\mathrm{\nu }\right)\ $ is given by Eq. \eqref{GrindEQ__22_}. Also in the large-\textit{$\nu$} region, the metric scale factor \textit{h${}_{R}$} can be transformed to a form where the sum is in denominator by the same approach as used for the small-\textit{$\nu$} region (see (D41)-(D43) in \textbf{Supplement D}):
\begin{equation} \label{GrindEQ__86_} 
h_R\mathrm{=\ \ }{\left[\mathrm{1-}\mu \sum^{\mathrm{\infty }}_{k\mathrm{=0}}{\left(\genfrac{}{}{0pt}{}{\frac{\mu }{\mathrm{1+}\mu }k-1}{k}\right)\frac{{\left(W^{\mathrm{-}\mathrm{\ }\frac{\mathrm{2}}{\mathrm{1+}\mathrm{\mu }}}\right)}^k}{\frac{\mu }{\mathrm{1+}\mu }k-1}\ }\right]}^{\mathrm{-}\mathrm{\ }\frac{\mathrm{1}}{\mathrm{2}}}\mathrm{\ \ } .   
\end{equation} 

\begin{figure}
	\includegraphics[width=\linewidth]{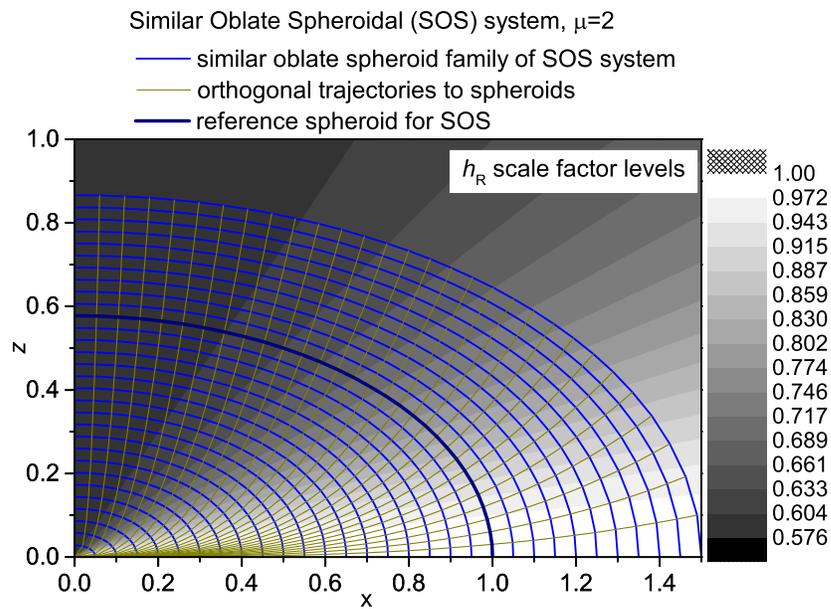}
	\caption{The metric scale factor \textit{h}${}_{R}$ levels in one quadrant of \textit{x}-\textit{z} plane for SOS coordinate system with \textit{µ}=2. The coordinate system lines (the same as in Figure \ref{fig:Fig1}) are overlapped here for a better comparison.}
	\label{fig:Fig3}
\end{figure}

For better imagination, 2D map of the \textit{h${}_{R}$} metric scale factor levels in \textit{x}-\textit{z} plane is shown in \textbf{Figure \ref{fig:Fig3}} for SOS system with \textit{µ}=2.

On the reference spheroid with the equatorial radius \textit{R}${}_{0}$, the metric scale factor \textit{h}${}_{R}$ can be even derived in a closed form (see \textbf{Supplement D,} Eq. (D65)):
\begin{equation} \label{GrindEQ__87_} 
h_{R0}\mathrm{=}\ \ \frac{1}{\sqrt{1+\mu {\mathrm{\ sin}}^{\mathrm{2}}\mathrm{\nu }}}\ .     
\end{equation} 
The derivation also leads to new binomial identities (D66), (D67) (see \textbf{Supplement D}).

When the convergence limits are calculated for the metric scale factor formulas \eqref{GrindEQ__82_} and \eqref{GrindEQ__85_} in the small-$\mathrm{\nu}$ region and in the large-$\mathrm{\nu}$ region, it is found that both regions have a common border line identical with the one defined in \eqref{GrindEQ__37_}. That means that the metric-scale-factor-defining power series have the same convergence interface between large- and small-\textit{$\nu$} regions as the power series for the calculation of the coordinates itself.

An important region at which the metric scale factor should be inspected is the abovementioned border between the small-\textit{$\nu$} region and the large-\textit{$\nu$} region (see \textbf{Figure \ref{fig:Fig2}}), where the power series in the formulas \eqref{GrindEQ__82_} and \eqref{GrindEQ__85_} may diverge, as discussed in the previous sections. It can be seen that the metric scale factor is a constant for constant value of \textit{W}, regardless the underlying values of \textit{R} and \textit{$\nu$}. Therefore, \textit{h}${}_{R}$ is constant on straight lines going through the origin (in 2D plot with generating ellipse; these constant regions are cones in 3D). Therefore, the border between the small-\textit{$\nu$} region and the large-\textit{$\nu$} regions also have a constant value of the metric scale factor, \textit{h}${}_{RB}$. It thus suffices to find this value once for ever for the particular \textit{µ} on one point on the border. The easiest is to find it on the intersection of the reference ellipse with the border line (\textit{h}${}_{R0B}$), and then the metric scale factor is determined on the whole border line as \textit{h}${}_{RB}$=\textit{h}${}_{R0B}$. For example, it was calculated in the previous sections that, for \textit{µ}=2 (i.e. for the ellipse used in \textbf{Figure \ref{fig:Fig2}} and for the corresponding generated spheroid), \textit{$\nu$}${}_{0B}$=0.331446717 rad (i.e. 18.99049802$\mathrm{{}^\circ}$), and \textit{h}${}_{RB}$ is thus -- using \eqref{GrindEQ__87_} -- equal to 0.908421069. Similarly, for \textit{µ}=0.006739496742 value, corresponding to the reference spheroid of the Earth, \textit{$\nu$}${}_{0B}$=0.77415419 rad (i.e. 44.35576776$\mathrm{{}^\circ}$) and \textit{h}${}_{RB}$ is thus\textit{ }equal to 0.9983570697.

For given \textit{µ}, we thus have the metric scale factor \textit{h}${}_{R}$ determined in the whole quadrant.

\noindent 
\subsection{ \textit{h}${}_{\boldsymbol{\nu }}$ metric scale factor}

For the derivation of \textit{h${}_{\nu }$}, Eq. \eqref{GrindEQ__70_} and \eqref{GrindEQ__71_} are employed. When the derivatives are input to the basic equation \eqref{GrindEQ__16_}, we obtain -- after rather lengthy simplification described in \textbf{Supplement D} -- the following formulas for the small-\textit{$\nu$} region: 

\begin{equation} \label{GrindEQ__88_} 
	\begin{aligned} 
&h_{\nu }\mathrm{=}\frac{R}{\sqrt{1+\mu }}\ \frac{\partial W}{\partial \nu }\sqrt{\sum^{\mathrm{\infty }}_{k\mathrm{=0}}{\left(\genfrac{}{}{0pt}{}{\left(1+\mu \right)\left(k+1\right)}{k}\right){\left({-W}^{\mathrm{2}}\right)}^k}}  \\ 
&=\frac{R}{\sqrt{1+\mu }}\ \frac{\partial W}{\partial \nu }\sqrt{\sum^{\mathrm{\infty }}_{k\mathrm{=0}}{\left(\genfrac{}{}{0pt}{}{-\mu \left(k+1\right)-2}{k}\right){\left(W^{\mathrm{2}}\right)}^k}}\  ,     
	\end{aligned} 
\end{equation} 
where the terms containing \textit{W} are determined by Eq. \eqref{GrindEQ__74_}. Two forms of the result are shown, one or the other can be more suitable in a particular case.

In the large-\textit{$\nu$} region, the simplification described in \textbf{Supplement D} leads to the following form of the metric scale factor \textit{h${}_{\nu }$} 

\begin{equation} \label{GrindEQ__89_} 
	\begin{aligned} 
&h_{\nu }=\frac{R}{1+\mu }W^{\mathrm{-}\mathrm{\ }\frac{\mathrm{2+}\mathrm{\mu }}{\mathrm{1+}\mathrm{\mu }}}\mathrm{\ }\frac{\partial W}{\partial \nu }\sqrt{\sum^{\mathrm{\infty }}_{k\mathrm{=0}}{\left(\genfrac{}{}{0pt}{}{{{\frac{1}{1+\mu }}}\left(k+1\right)}{k}\right){\left(\mathrm{-}W^{\mathrm{-}\mathrm{\ }\frac{\mathrm{2}}{\mathrm{1+}\mathrm{\mu }}}\right)}^k}}  \\ 
&=\frac{R}{1+\mu }W^{\mathrm{-}\mathrm{\ }\frac{\mathrm{2+}\mathrm{\mu }}{\mathrm{1+}\mathrm{\mu }}}\mathrm{\ }\frac{\partial W}{\partial \nu }\sqrt{\sum^{\mathrm{\infty }}_{k\mathrm{=0}}{\left(\genfrac{}{}{0pt}{}{{{\frac{\mu }{1+\mu }}}k-{{\frac{2+\mu }{1+\mu }}}}{k}\right){\left(W^{\mathrm{-}\mathrm{\ }\frac{\mathrm{2}}{\mathrm{1+}\mathrm{\mu }}}\right)}^k}}\ ,     
	\end{aligned} 
\end{equation}

\noindent which can be used for any \textit{µ $>$}--1. It should be also noted, that for $\mathrm{\nu}$=$\mathrm{\pi}$/2 (polar axis), substitutions of \textit{W}-containing terms according to \eqref{GrindEQ__74_} into \eqref{GrindEQ__89_} leads to a finite \textit{h${}_{\nu }$} equal to $R\ {\left({R}/{R_0}\right)}^{\mathrm{-}\mathrm{\ }\frac{\mu }{\mathrm{1+}\mathrm{\mu }}}\ \ \ $.

\begin{figure}
	\includegraphics[width=\linewidth]{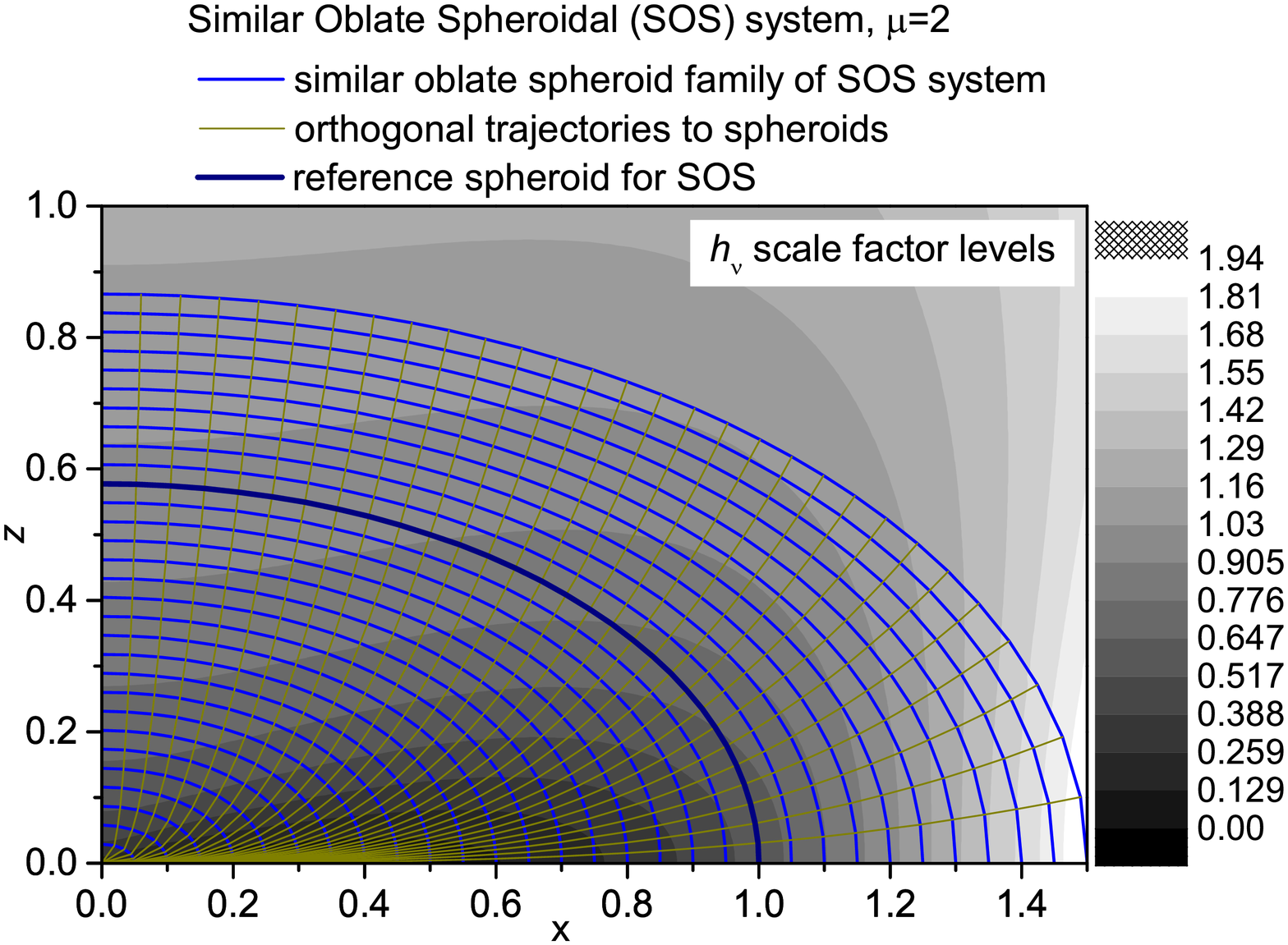}
	\caption{The metric scale factor \textit{h}${}_{\mathrm{\nu }}$ levels in one quadrant of \textit{x}-\textit{z} plane for SOS coordinate system with \textit{µ}=2. The coordinate system lines (the same as in Figure \ref{fig:Fig1}) are overlapped here for a better comparison.}
	\label{fig:Fig4}
\end{figure}

The convergence limits for the \textit{h${}_{\nu }$} metric-scale-factor defining power series are the same as for \textit{h${}_{R}$} and for coordinates itself. 2D map of the \textit{h${}_{\nu }$} metric scale factor levels in \textit{x}-\textit{z} plane is shown in \textbf{Figure \ref{fig:Fig4}} for SOS system with \textit{µ}=2.

On the reference spheroid with the equatorial radius \textit{R}${}_{0}$, the metric scale factor \textit{h}${}_{\mathrm{\nu }}$ can be derived in a closed form (see \textbf{Supplement D,} Eq. (D76)):
\begin{equation} \label{GrindEQ__90_} 
h_{\nu 0}\mathrm{=}R_0\ \sqrt{\frac{\mathrm{1+}\mathrm{\mu }\mathrm{\ }{\mathrm{sin}}^{\mathrm{2}}\mathrm{\nu }}{1+\mu }}\  .     
\end{equation} 

An important region at which the metric scale factor should be inspected is the border between the small-\textit{$\nu$} region and the large-\textit{$\nu$} region (see \textbf{Figure \ref{fig:Fig2}}), where the power series in the formulas \eqref{GrindEQ__88_} and \eqref{GrindEQ__89_} may diverge, as discussed in the previous sections.

It can be seen that the power series sums in metric scale factor formulas \eqref{GrindEQ__64_} and \eqref{GrindEQ__65_} are constant for constant value of \textit{W}, regardless the underlying values of \textit{R} and \textit{$\nu$}. Therefore, these power series sums are constant on straight lines going through the origin (in 2D plot with generating ellipse; these constant regions are cones in 3D).

Although the power series may diverge at the border line between the small-\textit{$\nu$} region and the large-\textit{$\nu$} regions, the metric scale factor has to have certain value there and thus the power series sum has to be exchanged by a certain value valid for the border line. It suffices to find this value once for ever for the particular \textit{µ} on one point on the border. The easiest is to find it on the intersection of the reference ellipse with the border line, which was in fact done in the previous paragraphs. When comparing \eqref{GrindEQ__88_} and \eqref{GrindEQ__90_} with help of \eqref{GrindEQ__74_}, it follows that the power series sum has to be exchanged on the border of the small-\textit{$\nu$} region by
\begin{equation} \label{GrindEQ__91_} 
\sqrt{\sum^{\mathrm{\infty }}_{k\mathrm{=0}}{\left(\genfrac{}{}{0pt}{}{\left(1+\mu \right)k+\left(1+\mu \right)}{k}\right){\left(\mathrm{-}W^2_{border}\right)}^k}}=\frac{{\mathrm{cos}}^{2+\mu }{\mathrm{\nu }}_{\mathrm{0B}}}{\sqrt{1+\mu \ {\mathrm{sin}}^{\mathrm{2}}{\mathrm{\nu }}_{\mathrm{0B}}}}\ \ ,   
\end{equation} 
Then, the metric scale factor anywhere on the border, i.e. for such \textit{R} and \textit{$\nu$} that
\begin{equation} \label{GrindEQ__92_} 
\mathrm{\ \ }{\left(\frac{R}{\mathrm{\ }R_0}\right)}^{\mu }\frac{{\mathrm{sin}}\mathrm{\nu }}{{\mathrm{co}\mathrm{s}}^{\mathrm{1+}\mathrm{\mu }}\mathrm{\ }\mathrm{\nu }}\mathrm{=}W_{border}\left(R,\nu \right)\mathrm{=}\sqrt{\frac{{\mu }^{\mu }}{{\left(\mathrm{1+}\mu \right)}^{\mathrm{1+}\mu }}}\ ,    
\end{equation} 
is -- by inserting \eqref{GrindEQ__91_} to \eqref{GrindEQ__88_} -- evaluated as

\begin{equation} \label{GrindEQ__93_} 
	\begin{aligned} 
&h_{\nu \mathrm{B}}\mathrm{=}\frac{R}{\sqrt{1+\mu }}\ \frac{\partial W}{\partial \nu }\frac{{\mathrm{cos}}^{2+\mu }{\mathrm{\nu }}_{\mathrm{0B}}}{\sqrt{1+\mu \ {\mathrm{sin}}^{\mathrm{2}}{\mathrm{\nu }}_{\mathrm{0B}}}}=\frac{R}{\sqrt{1+\mu }}{\left(\frac{R}{\mathrm{\ }R_0}\right)}^{\mu }\frac{\mathrm{1+}\mathrm{\mu }\mathrm{\ }{\mathrm{sin}}^{\mathrm{2}}\mathrm{\nu }}{{\mathrm{\ cos}}^{\mathrm{2+}\mathrm{\mu }}\mathrm{\ }\mathrm{\nu }}c_{\mathrm{\nu }\mathrm{B}}  \\ 
&\mathrm{\ \ \ where\ \ \ }c_{\mathrm{\nu }\mathrm{B}}\mathrm{=}c_{\mathrm{\nu }\mathrm{B}}\left(\mu \right)\mathrm{=}\frac{{\mathrm{cos}}^{2+\mu }{\mathrm{\nu }}_{\mathrm{0B}}}{\sqrt{1+\mu \ {\mathrm{sin}}^{\mathrm{2}}{\mathrm{\nu }}_{\mathrm{0B}}}}\ .  
	\end{aligned} 
\end{equation} 

\noindent For example, it was calculated in the previous sections that, for \textit{µ}=2 (i.e. for the ellipse used in \textbf{Figure \ref{fig:Fig2}} and for the corresponding generated spheroid), \textit{$\nu$}${}_{0B}$=0.331446717 rad (i.e. 18.99049802$\mathrm{{}^\circ}$), and \textit{c}${}_{\mathrm{\nu }B}$ is -- using \eqref{GrindEQ__93_} -- thus equal to 0.7262173699. Similarly, for \textit{µ}=0.006739496742, i.e. the value corresponding to the reference spheroid of the Earth, \textit{$\nu$}${}_{0B}$=0.77415419 rad (i.e. 44.35576776$\mathrm{{}^\circ}$) and \textit{c}${}_{\mathrm{\nu }B}$ is thus\textit{ }equal to 0.5092504742.

For given \textit{µ}, we thus have the metric scale factor \textit{h}${}_{\mathrm{\nu }}$ determined by analytical formulas in the whole quadrant.

\noindent 
\section{ Jacobian determinant}

Differential operations (gradient, divergence, curl, Laplacian) employ Jacobian determinant \textit{J} defined -- in the particular case of the SOS coordinates -- as  ${J=h_Rh_{\nu }h}_{\lambda }$ . Therefore, deriving Jacobian is very useful, similarly as the squares of the scale factor $h^2_R,\ {\ h}^2_{\nu },\ \ h^2_{\lambda }$ . The later are trivial, but the simplification of the Jacobian determinant is not, and is therefore done in \textbf{Supplement E}. The results for SOS coordinate system are listed in this section. 

For the small-\textit{$\nu$} region, the Jacobian is
\begin{equation} \label{GrindEQ__94_} 
\ {J=h_Rh_{\nu }h}_{\lambda }\mathrm{=}\frac{R^2}{\sqrt{1+\mu }}\ \frac{\partial W}{\partial \nu }\sum^{\mathrm{\infty }}_{k\mathrm{=0}}{\left(\genfrac{}{}{0pt}{}{\mathrm{-}\mu \left(k+{{\frac{1}{2}}}\right)\ -\ {{\frac{3}{2}}}}{k}\right){\left(W^{\mathrm{2}}\right)}^k}\ ,    
\end{equation} 
while for the large-\textit{$\nu$} region it reads
\begin{equation} \label{GrindEQ__95_} 
\ \ J=h_Rh_{\nu }h_{\lambda }\mathrm{=}\frac{{R^2W}^{\mathrm{-}\mathrm{\ }\frac{\mathrm{3+}\mathrm{\mu }}{\mathrm{1+}\mathrm{\mu }}}}{{\left(1+\mu \right)}^{{{\frac{3}{2}}}}}\ \frac{\partial W}{\partial \nu }\sum^{\mathrm{\infty }}_{k\mathrm{=0}}{\left(\genfrac{}{}{0pt}{}{\ \ {{\frac{\mu }{1+\mu }}}\left(k-{{\frac{1}{2}}}\right)\ -\ {{\frac{{\mathrm{3}}/{\mathrm{2}}}{1+\mu }}}}{k}\right){\left(W^{\mathrm{-}\mathrm{\ }\frac{\mathrm{2}}{\mathrm{1+}\mathrm{\mu }}}\right)}^k}\mathrm{\ } .   
\end{equation} 

As it frequently appears in differential operators, also the Jacobian divided by the square of the \textit{h${}_{R}$} scale factor was derived (see \textbf{Supplement E}, Eqs. (E15) and (E20)). This function is needed e.g. in the case of Laplacian calculation in SOS coordinates. In the small-$\mathrm{\nu}$ region, it equals to
\begin{equation} \label{GrindEQ__96_} 
\frac{J}{h^2_R}\mathrm{=}\frac{R^2}{\sqrt{1+\mu }}\ \frac{\partial W}{\partial \nu }\ \sum^{\mathrm{\infty }}_{k\mathrm{=0}}{\frac{{{\frac{\mathrm{\mu }\mathrm{+3}}{\mathrm{2}}\mathrm{\ }}}}{\mathrm{-}\mu \left(k+{{\frac{1}{2}}}\right)-{{\frac{3}{2}}}}\left(\genfrac{}{}{0pt}{}{\mathrm{-}\mu \left(k+{{\frac{1}{2}}}\right)-{{\frac{3}{2}}}}{k}\right){\left(W^{\mathrm{2}}\right)}^k\ } ,   
\end{equation} 
and -- in the large-\textit{$\nu$} region -- it is
\begin{equation} \label{GrindEQ__97_} 
\frac{J}{h^2_R}\mathrm{=}\frac{R^2}{\sqrt{1+\mu }}W^{\mathrm{-}\mathrm{\ }\frac{\mathrm{3+}\mathrm{\mu }}{\mathrm{1+}\mathrm{\mu }}}\ \frac{\partial W}{\partial \nu }\sum^{\mathrm{\infty }}_{k\mathrm{=0}}{\frac{{{\frac{1}{\mathrm{2}}\mathrm{\ \ }\frac{\mathrm{\mu }\mathrm{+}\mathrm{3}}{\mathrm{\mu }\mathrm{+1}}}}}{{{\frac{1}{\mathrm{2}}\mathrm{\ \ }\frac{\mathrm{\mu }\mathrm{+3}}{\mathrm{1+}\mathrm{\mu }}\mathrm{\ +\ }\frac{\mathrm{\mu }}{\mathrm{1+}\mathrm{\mu }}}}k}\left(\genfrac{}{}{0pt}{}{{{\frac{1}{\mathrm{2}}\mathrm{\ \ }\frac{\mathrm{\mu }\mathrm{+3}}{\mathrm{1+}\mathrm{\mu }}\mathrm{\ +\ }\frac{\mathrm{\mu }}{\mathrm{1+}\mathrm{\mu }}}}k}{k}\right){\left(W^{\mathrm{-}\mathrm{\ }\frac{\mathrm{2}}{\mathrm{1+}\mathrm{\mu }}}\right)}^k\ } .  
\end{equation}

\noindent 
\section{ Components of the advection term in the momentum equation}

\noindent In order to write the components of the momentum equation of atmosphere modelling in the curvilinear orthogonal system, the decomposition of the velocity advection term 
\begin{equation} \label{GrindEQ__98_} 
\mathrm{-}\left(\boldsymbol{\mathrm{u}}\boldsymbol{\mathrm{\ }}\mathrm{\textrm{·}\ }\mathrm{\nabla }\right)\boldsymbol{\mathrm{u}}\boldsymbol{\mathrm{\ }}\mathrm{=\ }\boldsymbol{\mathrm{u}}\boldsymbol{\mathrm{\ }}\mathrm{\times \ }\left(\mathrm{\nabla }\mathrm{\ \times }\boldsymbol{\mathrm{u}}\right)\mathrm{-}\mathrm{\nabla }\left(\frac{{\boldsymbol{\mathrm{u}}}^{\mathrm{2}}}{\mathrm{2}}\right) 
\end{equation} 
into its components is required. The components of the advection term (see \citet{Whi2008}, \citet{WI2011}, section 3, based on \citet{MoF1953}, p. 33) are, for SOS coordinates,
\[\mathrm{-}{\left[\left(\boldsymbol{\mathrm{u}}\boldsymbol{\mathrm{\ }}\mathrm{\textrm{·}\ }\mathrm{\nabla }\right)\boldsymbol{\mathrm{u}}\right]}_R\boldsymbol{\mathrm{\ }}\mathrm{=}\frac{u_{\nu }}{h_Rh_{\nu }}\left(u_{\nu }\frac{\partial h_{\nu }}{\partial R}-u_R\frac{\partial h_R}{\partial \nu }\right)\mathrm{+}\frac{u_{\lambda }}{h_Rh_{\lambda }}\left(u_{\lambda }\frac{\partial h_{\lambda }}{\partial R}-u_R\frac{\partial h_R}{\partial \lambda }\right)\mathrm{\ \ }\mathrm{\ -}\left(\boldsymbol{\mathrm{u}}\boldsymbol{\mathrm{\ }}\mathrm{\textrm{·}\ }\mathrm{\nabla }\right)u_R\mathrm{\ }\] 
\[\mathrm{-}{\left[\left(\boldsymbol{\mathrm{u}}\boldsymbol{\mathrm{\ }}\mathrm{\textrm{·}\ }\mathrm{\nabla }\right)\boldsymbol{\mathrm{u}}\right]}_{\nu }\boldsymbol{\mathrm{\ }}\mathrm{=}\frac{u_{\lambda }}{h_{\nu }h_{\lambda }}\left(u_{\lambda }\frac{\partial h_{\lambda }}{\partial \nu }-u_{\nu }\frac{\partial h_{\nu }}{\partial \lambda }\right)\mathrm{+}\frac{u_R}{h_{\nu }h_R}\left(u_R\frac{\partial h_R}{\partial \nu }-u_{\nu }\frac{\partial h_{\nu }}{\partial R}\right)\mathrm{\ \ \ -}\left(\boldsymbol{\mathrm{u}}\boldsymbol{\mathrm{\ }}\mathrm{\textrm{·}\ }\mathrm{\nabla }\right)u_{\nu }\mathrm{\ }\] 
\begin{equation} \label{GrindEQ__99_} 
\mathrm{-}{\left[\left(\boldsymbol{\mathrm{u}}\boldsymbol{\mathrm{\ }}\mathrm{\textrm{·}\ }\mathrm{\nabla }\right)\boldsymbol{\mathrm{u}}\right]}_{\lambda }\boldsymbol{\mathrm{\ }}\mathrm{=}\frac{u_R}{h_{\lambda }h_R}\left(u_R\frac{\partial h_R}{\partial \lambda }-u_{\lambda }\frac{\partial h_{\lambda }}{\partial R}\right)\mathrm{+}\frac{u_{\nu }}{h_{\lambda }h_{\nu }}\left(u_{\nu }\frac{\partial h_{\nu }}{\partial \lambda }-u_{\lambda }\frac{\partial h_{\lambda }}{\partial \nu }\right)\mathrm{\ \ \ -}\left(\boldsymbol{\mathrm{u}}\boldsymbol{\mathrm{\ }}\mathrm{\textrm{·}\ }\mathrm{\nabla }\right)u_{\lambda }\mathrm{\ }.   
\end{equation} 
As the partial derivatives of the metric scale factors with respect to \textit{$\lambda$} are zero, the expressions simplify to
\[\mathrm{-}{\left[\left(\boldsymbol{\mathrm{u}}\boldsymbol{\mathrm{\ }}\mathrm{\textrm{·}\ }\mathrm{\nabla }\right)\boldsymbol{\mathrm{u}}\right]}_R\boldsymbol{\mathrm{\ }}\mathrm{=}\frac{u_{\nu }}{h_Rh_{\nu }}\left(u_{\nu }\frac{\partial h_{\nu }}{\partial R}-u_R\frac{\partial h_R}{\partial \nu }\right)\mathrm{+}\frac{u_{\lambda }}{h_Rh_{\lambda }}u_{\lambda }\frac{\partial h_{\lambda }}{\partial R}\mathrm{\ \ \ -}\left(\boldsymbol{\mathrm{u}}\boldsymbol{\mathrm{\ }}\mathrm{\textrm{·}\ }\mathrm{\nabla }\right)u_R\mathrm{\ }\] 
\[\mathrm{-}{\left[\left(\boldsymbol{\mathrm{u}}\boldsymbol{\mathrm{\ }}\mathrm{\textrm{·}\ }\mathrm{\nabla }\right)\boldsymbol{\mathrm{u}}\right]}_{\nu }\boldsymbol{\mathrm{\ }}\mathrm{=}\frac{u_{\lambda }}{h_{\nu }h_{\lambda }}u_{\lambda }\frac{\partial h_{\lambda }}{\partial \nu }\mathrm{+}\frac{u_R}{h_{\nu }h_R}\left(u_R\frac{\partial h_R}{\partial \nu }-u_{\nu }\frac{\partial h_{\nu }}{\partial R}\right)\mathrm{\ \ \ -}\left(\boldsymbol{\mathrm{u}}\boldsymbol{\mathrm{\ }}\mathrm{\textrm{·}\ }\mathrm{\nabla }\right)u_{\nu }\mathrm{\ }\] 
\begin{equation} \label{GrindEQ__100_} 
\mathrm{-}{\left[\left(\boldsymbol{\mathrm{u}}\boldsymbol{\mathrm{\ }}\mathrm{\textrm{·}\ }\mathrm{\nabla }\right)\boldsymbol{\mathrm{u}}\right]}_{\lambda }\boldsymbol{\mathrm{\ }}\mathrm{=}-\frac{u_R}{h_{\lambda }h_R}u_{\lambda }\frac{\partial h_{\lambda }}{\partial R}-\frac{u_{\nu }}{h_{\lambda }h_{\nu }}u_{\lambda }\frac{\partial h_{\lambda }}{\partial \nu }\mathrm{\ \ \ -}\left(\boldsymbol{\mathrm{u}}\boldsymbol{\mathrm{\ }}\mathrm{\textrm{·}\ }\mathrm{\nabla }\right)u_{\lambda }\mathrm{\ } .   
\end{equation} 
It can be seen that the individual terms contain derivatives of the metric scale factors. There is eight terms containing such derivatives to be determined. Nevertheless, the unique coefficients in these terms containing only metric scale factors and their derivative are only four:
\begin{equation} \label{GrindEQ__101_} 
\ \ \frac{1}{h_Rh_{\lambda }}\frac{\partial h_{\lambda }}{\partial R}\ ,\ \ \ \frac{1}{{h_{\nu }h}_{\lambda }}\frac{\partial h_{\lambda }}{\partial \nu }\ ,\ \ \ \frac{1}{h_Rh_{\nu }}\frac{\partial h_{\nu }}{\partial R}\mathrm{\ }\ ,\ \ \ \frac{1}{{h_Rh}_{\nu }}\frac{\partial h_R}{\partial \nu } .    
\end{equation} 
It would be certainly useful to avoid the derivatives and to express these terms with the help of the already derived expressions for \textit{h}${}_{R}$, \textit{h}${}_{\mathrm{\nu }}$ and \textit{h}${}_{\mathrm{\lambda }}$. This is done in \textbf{Supplement F} where the previously derived relations are used to calculate the terms listed in \eqref{GrindEQ__101_}. The obtained relations for coefficients containing derivatives of the metric scale factors are listed here:
\[\frac{1}{h_Rh_{\lambda }}\frac{\partial h_{\lambda }}{\partial R}=\frac{h_R}{R}\] 
\[\frac{1}{{h_{\nu }h}_{\lambda }}\frac{\partial h_{\lambda }}{\partial \nu }=-\sqrt{\frac{1}{h^2_{\lambda }}-\frac{h^2_R}{R^2}}\mathrm{\ }\] 
\[\frac{1}{h_Rh_{\nu }}\frac{\partial h_{\nu }}{\partial R}=\frac{1}{R}\left[\left(\mu +2\right)h_R-\frac{1}{h_R}-\frac{\left(1+\mu \right){\mu }^2}{R^2}{\left(\frac{{\mathrm{sin} \nu \ }{\mathrm{cos} \nu \ }}{1+\mu {{\mathrm{sin}}^{\mathrm{2}} \nu \ }}\right)}^2h_Rh^2_{\nu }\right]\] 
\begin{equation} \label{GrindEQ__102_} 
\frac{1}{h_Rh_{\nu }}\frac{\partial h_R}{\partial \nu }=\mathrm{-}\frac{\mu \left(1+\mu \right)}{R^2}\frac{{\mathrm{sin} \nu \ }{\mathrm{cos} \nu \ }}{1+\mu {{\mathrm{sin}}^{\mathrm{2}} \nu \ }}h^2_Rh_{\nu } 
\end{equation} 
The relations are valid in both the small- and the large-\textit{$\nu$} region when proper expressions \textit{h}${}_{R}$, \textit{h}${}_{\mathrm{\nu }}$ and \textit{h}${}_{\mathrm{\lambda }}$ in the given region are used.

\noindent 

\noindent 
\section{ Gravitational potential and force in the interior of homogeneous oblate spheroid in SOS coordinates}

As a test and an illustration of the use of the derived formulas in the gravitational potential description, we transform the classical formula \citep{MMl1958} for the gravitational potential in the interior of a homogeneous oblate spheroid to the SOS coordinates, and also calculate constant force magnitude levels in these coordinates.

\citet{MMl1958} derived the equipotential surfaces in Cartesian coordinates in the form (\citet{MMl1958}, section 33) 

\begin{equation} \label{GrindEQ__103_} 
	\begin{aligned} 
&\frac{V-V_0}{\pi G\rho a^2c}\mathrm{=}\frac{x^2}{{\alpha }^2}+\frac{y^2}{{\alpha }^2}+\frac{z^2}{{\gamma }^2}  \\ 
&\mathrm{with}\ \ \ \ \ V_0=-\pi G\rho a^2c\int^{\infty }_0{\frac{\mathrm{d}s}{\left(a^2+s\right)\sqrt{c^2+s}}}=-\pi G\rho a^2c\ \frac{\pi -{\mathrm{2\ arcsin} \left({c}/{a}\right)\ }}{a\sqrt{1-{\left({c}/{a}\right)}^2}}\ .    
	\end{aligned} 
\end{equation} 
It should be noted that we changed the notation for the potential with respect to the original MacMillan one. Original MacMillan notation was increasing potential with the decreasing distance from the center, while we use here the standard present-day notation, i.e. the potential is decreasing towards the center, it is zero in the infinity, and negative elsewhere. Further, the original equation was written for a general ellipsoid while we simplified it for an oblate spheroid. Also, while MacMillan skipped gravitational constant \textit{G} in his derivations, we included it in our notation. In the previous formula \eqref{GrindEQ__103_}, \textit{V} is the potential, \textit{V}${}_{0}$ is its value (negative) in the center of the mass, and \textit{$\rho$} is the density (the same everywhere inside the homogeneous oblate spheroidal body). The letters \textit{a} and \textit{c} (\textit{a}$\mathrm{>}$\textit{c}) denote here the semi-axes (major and minor, respectively) of the oblate spheroid body. $\mathrm{\alpha}$ and $\mathrm{\gamma}$ are given by the integrals \citep{MMl1958} 

\[\frac{1}{{\alpha }^2}=\int^{\infty }_0{\frac{\mathrm{d}s}{\left(a^2+s\right)\sqrt{\left(a^2+s\right)\left(a^2+s\right)\left(c^2+s\right)}}}=\frac{-c\sqrt{a^2-c^2}+\frac{\pi }{2}a^2-a^2\mathrm{arctan}\frac{c}{\sqrt{a^2-c^2}}}{a^2{\left(a^2-c^2\right)}^{{{\frac{3}{2}}}}}\ , \] 

\begin{equation} \label{GrindEQ__104_} 
\frac{1}{{\gamma }^2}=\int^{\infty }_0{\frac{\mathrm{d}s}{\left(c^2+s\right)\sqrt{\left(a^2+s\right)\left(a^2+s\right)\left(c^2+s\right)}}}=\frac{2\sqrt{a^2-c^2}-\pi c+2c\ \mathrm{arctan}\frac{c}{\sqrt{a^2-c^2}}}{c{\left(a^2-c^2\right)}^{{{\frac{3}{2}}}}}\ .  
\end{equation} 
and the root of their ratio determines the ratio of the lengths of the semi-axes (major and minor) of the equipotential level surfaces inside the spheroidal body. In case of spheroids, these integrals can be solved in a closed form (in case of general ellipsoid not) and the solution is shown on the right side of \eqref{GrindEQ__104_}. Note also that (according to \citet{MMl1958})  ${{\alpha }^2}/{a^2}\mathrm{<}{{\gamma }^2}/{c^2}$, and the equipotential surfaces are thus not similar to the oblate spheroidal body -- they are more rounded.

The potential equation \eqref{GrindEQ__103_} written in Cartesian coordinates depends on all three coordinates: \textit{x}, \textit{y} and \textit{z}. Nevertheless, it can be written in SOS coordinates (\textit{R}, \textit{$\nu$}, \textit{$\lambda$}) in the form 

\begin{equation} \label{GrindEQ__105_} 
\frac{{\alpha }^2}{\pi G\rho a^2c}\left(V-V_0\right)\mathrm{=}R^2\ \ \ \ \ \ \ \ \ \mathrm{with}\ \ \ \ \ \ \ \ \ \ \ \ 1+\mu =\frac{{\alpha }^2}{{\gamma }^2}\ \ \ \ \ \ {{\stackrel{}{\Rightarrow}}}\ \ \ \ \frac{\gamma }{\alpha }={\left(1+\mu \right)}^{-{{\frac{1}{2}}}}\ .  
\end{equation} 

The potential
\begin{equation} \label{GrindEQ__106_} 
V\mathrm{=}\frac{\pi G\rho a^2c}{{\alpha }^2}R^2+V_0 
\end{equation} 
in SOS coordinates then depends, as expected, only on \textit{R }coordinate, not on the two others, \textit{$\nu$} and \textit{$\lambda$}, which can be of advantage for calculations. Further, its Laplacian have to fulfill Poisson equation (\citet{MMl1958}, section 69, \citet{Mor1990}, section 1.2), i.e.
\begin{equation} \label{GrindEQ__107_} 
\mathrm{\Delta }V\mathrm{=}4\pi G\rho  .      
\end{equation} 
When we write the Laplacian of the potential in SOS coordinates, we obtain (as the derivatives with respect to \textit{$\nu$} and \textit{$\lambda$} are zero)
\begin{equation} \label{GrindEQ__108_} 
\mathrm{\Delta }V\left(R\mathrm{,\ }\nu \mathrm{,\ }\lambda \right)\mathrm{=}\frac{\mathrm{1}}{J}\left[\frac{\partial }{\partial R}\left(\frac{J}{h^2_R}\frac{\partial V}{\partial R}\right)+0+0\right]=\frac{\mathrm{1}}{J}\frac{\partial }{\partial R}\left(\frac{J}{h^2_R}\frac{2\pi G{\rho a}^2c}{{\alpha }^2}R\right)\ .   
\end{equation} 

Using the formulas for the Jacobian over squared scale factor \textit{h${}_{R}$} \eqref{GrindEQ__96_} and for the Jacobian itself \eqref{GrindEQ__94_} in the small-$\mathrm{\nu}$ region, we can write the Laplacian of the potential in the form:

\begin{equation} \label{GrindEQ__109_} 
	\begin{aligned} 
&\mathrm{\Delta }V\left(R\mathrm{,\ }\nu \mathrm{,\ }\lambda \right)\mathrm{=}  \\ 
&\frac{\mathrm{1}}{J}\frac{\partial }{\partial R}\left\{\frac{2\pi G{\rho a}^2c}{{\alpha }^2}R\frac{R^2}{\sqrt{1+\mu }}\ \frac{R^{\mu }}{R^{\mu }_0}\frac{1+\mu {{\mathrm{sin}}^{\mathrm{2}} \nu \ }}{{{\mathrm{cos}}^{\mathrm{2+}\mathrm{\mu }} \nu \ }}\sum^{\mathrm{\infty }}_{k\mathrm{=0}}{\frac{{{\frac{\mathrm{\mu }\mathrm{+3}}{\mathrm{2}}\mathrm{\ }}}}{\mathrm{-}\mu \left(k+{{\frac{1}{2}}}\right)-{{\frac{3}{2}}}}\left(\genfrac{}{}{0pt}{}{\mathrm{-}\mu \left(k+{{\frac{1}{2}}}\right)-{{\frac{3}{2}}}}{k}\right){\left(W^{\mathrm{2}}\right)}^k}\right\}  \\ 
&=\frac{\mathrm{1}}{J}\left\{\frac{2\pi G{\rho a}^2c}{{\alpha }^2\sqrt{1+\mu }}\frac{1}{R^{\mu }_0}\frac{1+\mu {{\mathrm{sin}}^{\mathrm{2}} \nu }}{{{\mathrm{cos}}^{\mathrm{2+}\mathrm{\mu }} \nu }} \right. \\ 
& \left. \sum^{\mathrm{\infty }}_{k\mathrm{=0}}{\left(\genfrac{}{}{0pt}{}{\mathrm{-}\mu \left(k+{{\frac{1}{2}}}\right)-{{\frac{3}{2}}}}{k}\right)\frac{\mathrm{\mu }\mathrm{+3}}{\mathrm{2}\mu k+\mu +3}\left(\mathrm{2}\mu k+\mu +3\right)R^{2\mu k+\mu +2}\frac{1}{R^{2\mu k}_0}{\left(\frac{{\mathrm{sin} \nu }}{{{\mathrm{cos}}^{\mathrm{1+}\mathrm{\mu }} \nu }}\right)}^{\mathrm{2}k}}\right\}  \\ 
&=\frac{\frac{2\pi G{\rho a}^2c}{{\alpha }^2\sqrt{1+\mu }}\left(\mathrm{\mu }\mathrm{+3}\right)R^2\frac{\partial W}{\partial \nu }\sum^{\mathrm{\infty }}_{k\mathrm{=0}}{\left(\genfrac{}{}{0pt}{}{\mathrm{-}\mu \left(k+{{\frac{1}{2}}}\right)-{{\frac{3}{2}}}}{k}\right){\left(W^{\mathrm{2}}\right)}^k\ }}{\frac{R^2}{\sqrt{1+\mu }}\ \frac{\partial W}{\partial \nu }\sum^{\mathrm{\infty }}_{k\mathrm{=0}}{\left(\genfrac{}{}{0pt}{}{\mathrm{-}\mu \left(k+{{\frac{1}{2}}}\right)-{{\frac{3}{2}}}}{k}\right){\left(W^{\mathrm{2}}\right)}^k}}=\frac{2\pi G{\rho a}^2c}{{\alpha }^2}\left(\mathrm{\mu }\mathrm{+1+2}\right)  \\ 
&=2\pi G\rho a^2c\frac{1}{{\alpha }^2}\left(\frac{{\alpha }^2}{{\gamma }^2}+2\right)=2\pi G\rho a^2c\left(\frac{1}{{\gamma }^2}+\frac{2}{{\alpha }^2}\right)\ . 
	\end{aligned} 
\end{equation} 

\noindent In order to fulfill the Poisson equation \eqref{GrindEQ__107_},  $\left({1}/{{\gamma }^2}+{2}/{{\alpha }^2}\right)$ in the last expression in the above relation has to be equal to ${2}/{\left(a^2c\right)}$. This is exactly what follows from the summation using the evaluated integrals \eqref{GrindEQ__104_}.

We can also determine 1+\textit{µ} factor for the equipotential surfaces from the integrals in \eqref{GrindEQ__104_} using \eqref{GrindEQ__105_}. Then
\begin{equation} \label{GrindEQ__110_} 
1+\mu =\frac{{\alpha }^2}{{\gamma }^2}=-2\frac{\frac{1}{{c}/{a}}\sqrt{1-{\left({c}/{a}\right)}^2}-\mathrm{arcos}\left({c}/{a}\right)}{\left({c}/{a}\right)\sqrt{1-{\left({c}/{a}\right)}^2}-\mathrm{arcos}\left({c}/{a}\right)} 
\end{equation} 
and the parameter 1+\textit{µ} for the family of equipotential oblate spheroids can be calculated from the semi-axes ratio \textit{c}/\textit{a} of the original spheroidal body. Although it is not visible at once (L'Hopital's rule has to be used), the expression on the right side is at the limit when \textit{c} is approaching \textit{a} (i.e. for a spherical body) equal to 1, and \textit{µ}=0 in this case, as is expected for a sphere.

The force per unit mass (or acceleration) inside the homogeneous oblate spheroid can be calculated from the gravitational potential \eqref{GrindEQ__106_} by using the gradient operator:
\begin{equation} \label{GrindEQ__111_} 
\boldsymbol{F}\mathrm{=}\boldsymbol{\mathrm{\nabla }}V\mathrm{=}\frac{{\widehat{\boldsymbol{\mathrm{e}}}}_R}{h_R}\frac{\partial V}{\partial R}\mathrm{+}\frac{{\widehat{\boldsymbol{\mathrm{e}}}}_{\nu }}{h_{\nu }}\frac{\partial V}{\partial \nu }\mathrm{+}\frac{{\widehat{\boldsymbol{\mathrm{e}}}}_{\lambda }}{h_{\lambda }}\frac{\partial V}{\partial \lambda }\mathrm{=}\frac{{\widehat{\boldsymbol{\mathrm{e}}}}_R}{h_R}\frac{\partial V}{\partial R}+0+0\mathrm{=}{\widehat{\boldsymbol{\mathrm{e}}}}_R\mathrm{\ 2}\frac{\pi G\rho a^2c}{{\alpha }^2}\ \frac{R}{h_R}  ,  
\end{equation} 
where ${\widehat{\boldsymbol{\mathrm{e}}}}_R,\ {\widehat{\boldsymbol{\mathrm{e}}}}_{\nu }\mathrm{,\ }{\widehat{\boldsymbol{\mathrm{e}}}}_{\lambda }$ are the unit vectors in the direction of the individual SOS coordinates. In SOS coordinates, we advantageously used the fact that the potential depends on one variable only. The force magnitude 
\begin{equation} \label{GrindEQ__112_} 
\left|\boldsymbol{F}\right|\mathrm{=}\mathrm{\ 2}\frac{\pi G\rho a^2c}{{\alpha }^2}\ \frac{R}{h_R}\ ,     
\end{equation} 
however, is evidently having a component dependent on \textit{$\nu$}--coordinate in the used SOS coordinate system with the parameter \textit{µ}, as \textit{h}${}_{R}$ depends also on \textit{$\nu$}--coordinate through the \textit{W} parameter. 

Nevertheless, the force magnitude equi-surfaces could still be of an oblate\linebreak spheroidal shape; only these oblate spheroids could not be from the same family as the oblate spheroids for the equipotential surfaces. In such a case, in a new SOS coordinate system (\textit{R}${}_{F}$, \textit{$\nu$}${}_{F}$, \textit{$\lambda$}${}_{F}$) with the parameter \textit{µ}${}_{F}$, it would be possible to write for every \textit{R}${}_{F}$ inside the oblate spheroid body the force magnitude in the form 
\begin{equation} \label{GrindEQ__113_} 
\left|\boldsymbol{F}\right|\mathrm{=}\mathrm{\ 2}\frac{\pi G\rho a^2c}{{\alpha }^2}R_{\mathrm{F}}\  ,     
\end{equation} 
i.e. depending on \textit{R}${}_{F}$ only. To test this possibility (i.e. to fit potentially the new SOS coordinate system onto the equi-force-magnitude levels), we need a transformation relation for the \textit{R}${}_{F}$ coordinate. The derivation which can be found in \textbf{Supplement G} shows (see (G5)) that
\begin{equation} \label{GrindEQ__114_} 
R_F=R\sqrt{\frac{1+{\mu }_F}{\ 1+\mu }+\frac{{\mu }_F-\mu }{\ 1+\mu }\sum^{\infty }_{M=0}{\frac{1}{-\mu M-1}\left(\genfrac{}{}{0pt}{}{-\mu M-1}{M}\right){\left(W^2\right)}^M}} .  
\end{equation} 
in the small-\textit{$\nu$} region. Insertion of the coordinate \textit{R} from the transformation relation \eqref{GrindEQ__114_} and \textit{h}${}_{R}$ from the relation \eqref{GrindEQ__81_} to the equation \eqref{GrindEQ__112_} resuls in
\begin{equation} \label{GrindEQ__115_} 
\left|\boldsymbol{F}\right|\mathrm{=}\mathrm{\ 2}\frac{\pi G\rho a^2c}{{\alpha }^2}R_{\mathrm{F}}\frac{\sqrt{\left(\mathrm{1+}\mu \right)+\mu \sum^{\mathrm{\infty }}_{k\mathrm{=0}}{{{\frac{\mathrm{1}}{\mathrm{-}\mu k-1}}}\left(\genfrac{}{}{0pt}{}{-\mu k-1}{k}\right){\left(W^{\mathrm{2}}\right)}^k\ }}}{\sqrt{{{\frac{1+{\mu }_{\mathrm{F}}}{\mathrm{1+}\mu }}}\mathrm{+}{{\frac{{\mu }_{\mathrm{F}}\mathrm{-}\mu }{\mathrm{1+}\mu }}}\sum^{\mathrm{\infty }}_{M\mathrm{=0}}{{{\frac{\mathrm{1}}{\mathrm{-}\mu M-1}}}\left(\genfrac{}{}{0pt}{}{\mathrm{-}\mu M-1}{M}\right){\left(W^{\mathrm{2}}\right)}^M}}}\   .  
\end{equation} 

In order to be able to write the force magnitude relation in the form \eqref{GrindEQ__113_}, the ratio which includes the power series in the numerator and in the denominator on the right of \eqref{GrindEQ__115_} would need to be equal to one for every \textit{W}. It leads to the conditions 
\begin{equation} \label{GrindEQ__116_} 
\left(\mathrm{1+}\mu \right)\mathrm{=}\frac{1+{\mu }_{\mathrm{F}}}{\ 1+\mu }\mathrm{\ \ \ \ \ \ \ \ \ \ and\ \ \ \ \ \ \ \ \ \ }\mu \mathrm{=}\frac{{\mu }_{\mathrm{F}}\mathrm{-}\mu }{\ 1+\mu }\  ,     
\end{equation} 
which would have to be fulfilled at the same time. Simple algebraic elaboration leads in both cases to the same result 
\begin{equation} \label{GrindEQ__117_} 
1+{\mu }_{\mathrm{F}}\mathrm{=}{\left(1+\mu \right)}^2\ \ \ \ \  .     
\end{equation} 
It shows that the force magnitude can be really written in the form \eqref{GrindEQ__113_}. Therefore, a SOS coordinate system (\textit{R}${}_{F}$, \textit{$\nu$}${}_{F}$, \textit{$\lambda$}${}_{F}$) with the parameter \textit{µ}${}_{F}$ according to \eqref{GrindEQ__117_} condition does exist, in which the equi-force-magnitude levels can be expressed depending only on the SOS coordinate \textit{R}${}_{F}$. Then, the equi-force-magnitude levels are oblate spheroids. Note, that the condition \eqref{GrindEQ__117_} corresponds (according to \eqref{GrindEQ__105_}) to

\begin{equation} \label{GrindEQ__118_} 
1+{\mu }_{\mathrm{F}}=\frac{{\alpha }^4}{{\gamma }^4}  ,     
\end{equation} 
which is the same condition as the one which can be deduced from the classical derivation using Cartesian coordinates (\citet{MMl1958}, section 34).

\begin{figure}
	\includegraphics[width=\linewidth]{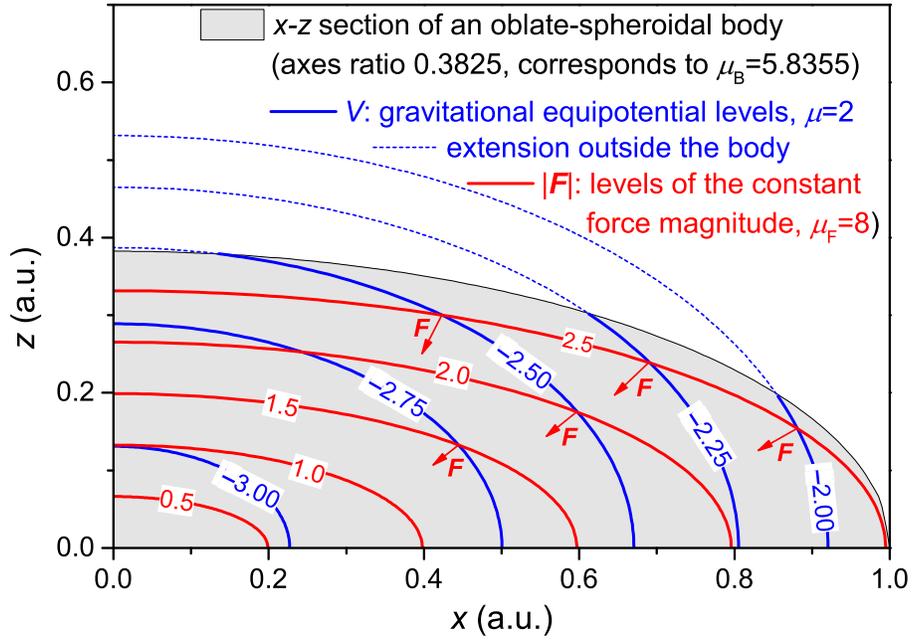}
	\caption{One quadrant showing \textit{x}-\textit{z} section of the homogeneous oblate-spheroidal body with the axes ratio 0.3825 (corresponds to \textit{$\mu$} parameter of the oblate body \textit{$\mu$}${}_{B}$=5.8355), corresponding gravitational equipotential levels (oblate spheroid family with \textit{$\mu$}=2), and the corresponding levels of the constant force magnitude (oblate spheroid family with \textit{µ}${}_{F}$=8). Also displayed is the extension of the interior equipotential levels outside the body (it has to be stressed that this extension does not correspond to the theoretical equipotential levels outside the oblate spheroidal body). Finally, the direction of the force vector F is illustrated at some points of the scheme.}
	\label{fig:Fig5}
\end{figure}

For illustration of the abovementioned formulas, \textbf{Figure \ref{fig:Fig5}} depicts the oblate\linebreak spheroidal body and its (mutually connected) interior levels of the oblate spheroidal shape: (\textit{i}) the equipotential surfaces and (\textit{ii}) the equi-force-magnitude levels. It can be seen that while the interior equipotential surfaces are more rounded than the oblate spheroidal body, the levels of the constant force magnitude are -- on the contrary -- more flattened than the oblate spheroidal body itself.

It also does worth to remind (as previously pointed out generally by \citet{Hof2018}) that the force \textbf{\textit{F}} is non-central (except of the equator and the polar axis), and that -- except the equatorial plane -- its vertical component exists even for a very much flattened oblate spheroid. In a galaxy approximated as an oblate spheroid, this force component acting on a star cannot be compensated by the centrifugal force caused by a simple rotation around the vertical axis of the galaxy \citep{Hof2017}. This fact has to have a consequence on the trajectories of stars in galaxies.

\noindent 

\noindent 
\section{ Numerical procedure}

FORTRAN 95 code was created for calculation of \textit{x}, \textit{z} Cartesian coordinates from the \textit{R} and \textit{$\nu$} SOS coordinates using power series with generalized binomial coefficients derived in this paper. The partial derivatives calculation is also included in the code. Finally, the metric scale factors \textit{h}${}_{R}$ and \textit{h${}_{\nu }$} are numerically computed as well. The code is attached to this article as \textbf{Supplement H}.

The code was written basically as the test of the correctness of the analytical formulas. With this code, the data for the figures presented in this paper were calculated. Further, the convergence of the power series was tested. It appeared that the convergence is unsatisfactory for \textit{µ} equal to zero in the vicinity of the border line between the small- and the large-\textit{$\nu$} regions. It is not a problem as the spherical coordinates can be used for \textit{µ}=0 instead. Above zero value of \textit{µ}, the convergence of the power series for the coordinates and the metric scale factors is satisfactory. It means that also for the Earth reference ellipsoid, which has rather small oblateness (\textit{$\mu$} = 0.006739496742), the code can be used. The convergence was tested up to \textit{µ}=169 (i.e. the semi-axes ratio 0.077), above which value the double-precision real-number limit (around 10${}^{300}$) of the FORTRAN code is exceeded for the parts of the large-\textit{$\nu$} region. Above this value, some more sophisticated numerical procedure would need to be used.

It was also found that the partial derivatives converge rather badly. Nevertheless (which is significantly more important), the scale factors itself converge very well except the vicinity of the border line between the small- and the large-\textit{$\nu$} regions. The very vicinity of the border line between the small- and the large-\textit{$\nu$} regions can be generally problematic for the power series convergence. Some smart, better converging solution could be searched for in the further work.

It is probably possible to use the procedure (with certain modifications) also for prolate spheroidal (\textit{µ} would fulfill the condition --1$\mathrm{<}$\textit{ µ} $\mathrm{<}$0 in such case) coordinates calculation. Nevertheless, this option was not tested in the presented work.

\noindent 
\section{ Conclusions}

Explicit analytical expressions for the Cartesian coordinates in terms of the curvilinear Similar Oblate Spheroidal coordinates (\textit{R}, \textit{$\nu$}, \textit{$\lambda$}) were derived in the form of infinite power series with generalized binomial coefficients (see Eq. \eqref{GrindEQ__52_}, \eqref{GrindEQ__53_}) on the basis of Lagrange Inversion Formula. On this ground, the partial derivatives with respect to the individual SOS coordinates were derived. It further enabled derivation of the scale factors \textit{h}${}_{R}$, \textit{h${}_{\nu }$}, \textit{h${}_{\lambda }$} (see Eq. \eqref{GrindEQ__80_}, \eqref{GrindEQ__81_}, \eqref{GrindEQ__82_}, \eqref{GrindEQ__85_}, \eqref{GrindEQ__88_}, \eqref{GrindEQ__89_}), which are necessary for differential operations using SOS coordinates. The solution had to be divided into two regions denoted the small-\textit{$\nu$} region and the large-\textit{$\nu$} region, in which different form of the power series is needed in order to ensure the power-series convergence. The convergence and analytic solution on the border line between the small-\textit{$\nu$} region and the large-\textit{$\nu$} region were discussed.

Although the derivation of the scale factors is lengthy, the result -- in the form of infinite power series with generalized binomial coefficients -- is surprisingly simple. As a by-product of the derivation, new binomial identities were found (see Suplement D, formulas (D17), (D66), (D67)). The analytic solution of the metric scale factors on the border line between the small-\textit{$\nu$} region and the large-\textit{$\nu$} region was found and discussed. 

The derived SOS-coordinates expressions for metric scale factors can be used in the analytical formulas containing differential operators. The terms containing derivatives of the metric scale factors in the velocity advection term of the momentum equation in SOS coordinate system were expressed by formulas using metric scale factors (see Eq. \eqref{GrindEQ__102_}). Further, Jacobian determinant \textit{h}${}_{R}$\textit{h${}_{\nu }$h${}_{\lambda }$} was derived (see Eq. \eqref{GrindEQ__94_}, \eqref{GrindEQ__95_}). Gravitational potential (see Eq. \eqref{GrindEQ__105_}, \eqref{GrindEQ__106_}) and force (see Eq. \eqref{GrindEQ__113_}, \eqref{GrindEQ__116_}) in the interior of homogeneous oblate spheroid in the SOS coordinates was derived as well. 

The received formulas were tested numerically using FORTRAN code, which is attached as a supplementary material to this article.

The SOS coordinate system can be advantageously used in cases when gravitational potential levels, gravity potential levels or density levels fit with, or can be well approximated by, similar oblate spheroids family. Then, the quantity would depend on one coordinate of the SOS system only, which lowers the dimension of the mathematical description of the related physical processes. E.g., it would lead to an essential simplification of differential equations.

The presented SOS coordinates solution is applicable for a broad variety of objects, ranging from the bodies with small oblateness (like the Earth with semi-axes ratio around 1/300 and corresponding \textit{$\mu$} = 0.006739496742), through elliptical galaxies (e.g., \textit{$\mu$} $\mathrm{\approx}$3) up to significantly flattened objects like disk galaxies (\textit{µ} values in hundreds). There is (untested) possibility that the derived formulas, or at least the used approach, can be with some modifications applied also for analytic derivation of similar prolate spheroidal coordinates with --1$\mathrm{<}$\textit{ µ} $\mathrm{<}$0.

The analytical solution for SOS coordinates provides potentially a new tool for modelling density, potential and acceleration inside or in the vicinity of oblate/linebreak spheroidal objects, like planets and galaxies. Modelling of the near-Earth geopotential for the use in atmosphere dynamics calculations is particularly to be mentioned. Further, models for an explanation of the rotational curves of galaxies could be possibly formulated with a help of the SOS coordinates.

Follow-up work in this direction could include testing if Laplace differential equation is separable in SOS coordinates. It is also clear that the convergence/divergence of the power series which form basis of the SOS to Cartesian coordinates transformation and metric scale factors calculation is ambiguous on one line in \textit{x}-\textit{z} plane (or on one conical surface in 3D). This point should be clarified by experienced mathematicians. Also, the same approach could be used highly probably for similar prolate spheroidal (SPS) coordinates. \\

\noindent \textbf{\textit{Acknowledgments.}}

\noindent The author acknowledges partial support from the Strategie AV21 programme No. 23 of the CAS. The author declares he has no conflict of interests.

\noindent \\

\noindent \textbf{\textit{Data Availability Statement.}}

\noindent This paper deals with the derivation of theoretical relations of SOS coordinate system. The only data created are those created by means of FORTRAN code attached in \textbf{Supplement H}, which can be easily reproduced.

\noindent 

\noindent 

\noindent

\end{document}